\pdfoutput=1
\documentclass[preprint]{imsart}

\usepackage{amsthm,amsmath}

\usepackage{subfigure,tikz}
\usetikzlibrary{backgrounds,automata}

\usepackage{natbib}
\usepackage{multirow}
\usepackage{color,soul}
\usepackage{arydshln}
\usepackage{bbold}
\usepackage{rotating}
\usepackage{multicol}

\usepackage{helvet}
\newcommand{\norm}[1]{\lvert #1 \rvert}
\newcommand{\twopartdef}[4]
{	\left\{
		\begin{array}{ll}
			#1 & \mbox{if } #2 \\
			#3 & \mbox{if } #4
		\end{array}
	\right.}

\usepackage{ upgreek }

 \usepackage{tikz}
 \def\firstcircle{(.5,.5) circle (0.8cm)}
  \def\secondcircle{(1,1) circle (.8cm)}
  \def\thirdcircle{(0,1) circle (.8cm)}
  \def\fourthcircle{(0,1) circle (.8cm)}
  \def\fifthcircle{(-.1,.9) circle (.5cm)}
\usetikzlibrary{shapes,backgrounds}
\usepackage{wrapfig}
\usepackage{lmodern}
\usepackage{mathtools}
\usepackage{blindtext}
\usepackage{graphicx}
\usepackage{caption}
\usepackage{subfigure}
\newcommand{\undertilde}[1]{\ensuremath{\mathord{\vtop{\ialign{##\crcr
  $\hfil\displaystyle{#1}\hfil$\crcr\noalign{\kern1.5pt\nointerlineskip}
  $\hfil\tilde{}\hfil$\crcr\noalign{\kern1.5pt}}}}}}
\usepackage[margin=1in]{geometry}
\usepackage[]{graphicx}\usepackage[]{color}

\newdimen\bibsep

\newdimen\bibhang
\begin{document}

\begin{frontmatter}

\title{Estimation of causal effects with multiple treatments: a review and new ideas}
\runtitle{Matching with multiple treatments}

\begin{aug}
\author{\fnms{Michael J.} \snm{Lopez}\thanksref{t1}\ead[label=e1]{mlopez1@skidmore.edu}},
\and
\author{\fnms{Roee} \snm{Gutman}\ead[label=e2]{roee\_gutman@brown.edu}}

\thankstext{t1}{MJ Lopez was supported by the National Institute for Health (IMSD Grant \#R25GM083270) and the National Institute on Aging (NIA grant \#F31AG046056)}
\thankstext{t2}{R Gutman was partially supported through a Patient-Centered Outcomes Research Institute (PCORI) Award ME-1403-12104. Disclaimer: All statements in this report, including its findings and conclusions, are solely those of the authors and do not necessarily represent the views of the Patient-Centered Outcomes Research Institute (PCORI), its Board of Governors or Methodology Committee.}

\runauthor{Lopez \& Gutman}

\affiliation{Skidmore College \printead{e1}}
\affiliation{Brown University \printead{e2}}

\address{Department of Mathematics, 
815 N. Broadway, Saratoga Springs, NY 12866 \printead{e1}.}

\address{Department of Biostatistics, Box G-S121-7, 
121 S. Main Street, Providence, RI 02912 \printead{e2}.}

\end{aug}

\begin{abstract} 

The propensity score is a common tool for estimating the causal effect of a binary treatment in observational data. In this setting, matching, subclassification, imputation, or inverse probability weighting on the propensity score can reduce the initial covariate bias between the treatment and control groups. With more than two treatment options, however, estimation of causal effects requires additional assumptions and techniques, the implementations of which have varied across disciplines. This paper reviews current methods, and it identifies and contrasts the treatment effects that each one estimates. Additionally, we propose possible matching techniques for use with multiple, nominal categorical treatments, and use simulations to show how such algorithms can yield improved covariate similarity between those in the matched sets, relative the pre-matched cohort. To sum, this manuscript provides a synopsis of how to notate and use causal methods for categorical treatments. 

\end{abstract}


\end{frontmatter}

\doublespacing

\section{Introduction} \label{sec21}

The primary goal of many scientific applications is to identify the causal effect of exposure $T \in \left\{ t_1, ..., t_Z\right\}$ on outcome $Y$. Randomized experiments are the gold standard for estimating a causal relationship, however, they are sometimes infeasible due to logistical, ethical, or financial considerations. Further, randomized experiments may not be as generalizable as observational studies due to the restricted population used in the experiments. 

When assignment to treatment is not randomized, those that receive one level of the treatment may differ from those that receive another with respect to covariates, $\boldsymbol{X}$, that may also influence the outcome. For example, in a study estimating the causal effects of neighborhood choice on employment, persons who live in deprived neighborhoods differ from those who live in privileged ones on a variety of characteristics, such as socioeconomic status and education levels \citep{hedman2012understanding}. As such, it may be difficult to distinguish between neighborhood effects and the differences between subjects which existed before they chose their neighborhoods. In such settings, establishing causes and effects requires more sophisticated statistical tools and additional assumptions. 

Methods such as matching \citep{dehejia2002propensity}, subclassification \citep{rosenbaum1984reducing}, weighting \citep{robins2000marginal}, and imputations \citep{gutman2015estimation} have been proposed to adjust for the differences in $\boldsymbol{X}$ across the exposure groups. These approaches attempt to obtain covariate balance across treatment groups, where balance refers to equality in the distributions of $\boldsymbol{X}$. By ensuring that the distribution of units receiving different treatments are similar on average, these methods attempt to reproduce a randomized trial, thus reducing the effects of treatment assignment bias on causal estimates. 

When $\boldsymbol{X}$ is a scalar, it is relatively straight-forward to perform matching \citep{rubin1976multivariate}. However, it is more complex to match, subclassify, or weight when $\boldsymbol{X}$ is composed of many covariates. With a binary treatment, matching, subclassification, weighting, and imputation using the propensity score have been proposed for estimating causal effects from observational studies with binary treatment \citep{rosenbaum1983central, stuart2010matching, gutman2015estimation}. Propensity score is defined as the probability of receiving the treatment conditional on a set of observed covariates. It has been shown in theory \citep{rubin1996matching} and practice \citep{d1998tutorial, caliendo2008some} that under certain assumptions, matching on propensity scores results in unbiased unit-level estimates of the treatment's causal effect \citep{rosenbaum1983central}.

Generalizations and applications of propensity score methods for multiple treatments, however, remain scattered in the literature, in large part because the advanced techniques are unfamiliar and inaccessible. Our first goal is to provide a unifying terminology that will enable researchers to coalesce and compare existing methods. Our second goal is to describe current methods for estimating causal effects with multiple treatments, with a specific focus on approaches for nominal categorical exposures (e.g., a comparison of pain-killers Motrin, Advil, and Tylenol). We contrast these methods' assumptions and define the causal effects they each attempt to estimate. In doing so, potential pitfalls in the commonly used practice of applying binary propensity score tools to multiple treatments are identified. 

Third, we explain the elevated importance of defining a common support region when studying multiple treatments, where differences in the implementation of certain approaches can vary the causal estimands as well as change the study population to which inference is generalizable. Our final goal is to provide a technique for generating matched sets when there are more than two treatments that addresses some of the pitfalls of the current methods, as well as to compare the performance of the new and previously proposed algorithms in balancing covariates' distributions using extensive simulation analysis. 

The remainder of Section \ref{sec21} introduces the notation and identifies existing causal methods for multiple treatments. Section \ref{VM} proposes a new algorithm for matching with multiple treatments. Section \ref{Sim} uses simulations to contrast the new and previously proposed approaches for generating well-matched subgroups of subjects. Section \ref{Disc} discusses and concludes with a set of practical recommendations.

\subsection{Notation for binary treatment} 

Our notation is based on the potential outcomes framework, originally proposed by Neyman for randomized based inference, and extended by Rubin to observational studies and Bayesian analysis, also known as the Rubin Causal Model (RCM) \citep{splawa1990application, rubin1975bayesian, holland1986statistics}. Let $Y_i$, $\boldsymbol{X_i}$, and $T_i$ be the observed outcome, set of covariates, and binary treatment assignment, respectively, for each subject \emph{i = 1,..., N}, with $N \leq \mathcal{N}$, where $\mathcal{N}$ is the population size which is possibly infinite. With $T_i \in \mathcal{T}$, let $\mathcal{T}$ be the treatment space. For a binary treatment, $\mathcal{T} = \left \{t_1,t_2\right \}$, and let $n_{t_1}$ and $n_{t_2}$ be the number of subjects receiving treatments $t_1$ and $t_2$, respectively. 

The RCM relies on the Stable Unit Treatment Value Assumption ($SUTVA$) to define the potential outcomes $Y_i(t_1)$ and $Y_i(t_2)$, which would have been observed had unit $i$ simultaneously received $t_1$ and $t_2$, respectively \citep{rubin1980comment}. SUTVA specifies no interference between subjects and no hidden treatment versions, entailing that the set of potential outcomes for each subject does not vary with the treatment assignment of others. Because each individual receives only one treatment at a specific point in time, only $Y_i(t_1)$ or $Y_i(t_2)$ is observed for each subject, which is known as the fundamental problem of causal inference \citep{holland1986statistics}. 

Two commonly used estimands for describing super-population effects are the population average treatment effect, $PATE_{t_1,t_2}$, and the population average treatment effect among those receiving $t_1$, $PATT_{t_1, t_2}$. \begin{eqnarray} 
	PATE_{t_1, t_2} &=&  E[Y_i(t_1)-Y_i(t_2)] \label{e1} \\
	PATT_{t_1, t_2} &=& E[Y_i(t_1)-Y_i(t_2)|T_i=t_1] \label{e2}\end{eqnarray}

Letting $I(T_i = t_1)$ be the indicator function for an individual receiving treatment $t_1$, $PATE_{t_1, t_2}$ and $PATT_{t_1, t_2}$ are generally approximated by the sample average treatment effects. 
\begin{eqnarray} 
	SATE_{t_1, t_2} &=& \frac{1}{N}\sum_{i=1}^N (Y_i(t_1)-Y_i(t_2)) \label{ate} \\
	SATT_{t_1, t_2} &=& \frac{1}{n_{t_1}}\sum_{i=1}^N (Y_i(t_1)-Y_i(t_2)) \times I(T_i = t_1) \label{att} \end{eqnarray}

Because only one of the potential outcomes is observed for every unit, an important piece of information to estimate (\ref{ate}) and (\ref{att}) is the assignment mechanism, $P(\boldsymbol{T}|\boldsymbol{Y(t_1)},\boldsymbol{Y(t_2)},\boldsymbol{X})$, where $\boldsymbol{T} = \left\{T_i \right\}$, $\boldsymbol{Y(t_1)} = \left\{Y_i(t_1) \right\}$, $\boldsymbol{Y(t_2)} = \left\{Y_i(t_2) \right\}$ and $\boldsymbol{X} = \left\{\boldsymbol{X_i} \right\}$  \citep{ImbensRubinBook}. Three commonly made restrictions of the assignment mechanism are individualistic, probabilistic, and unconfoundedness \citep{ImbensRubinBook}. In the super population, a random sample of $N$ units automatically results in an individualistic assignment mechanism. A super-population probabilistic assignment mechanism entails that \begin{eqnarray} 0 \ \textless \ f_{\boldsymbol{T} | \boldsymbol{Y(0)}, \boldsymbol{Y(1)}, \boldsymbol{X}} (t_1 | Y_i(0), Y_i(1), \boldsymbol{X_i}, \phi) \ \textless \ 1 \nonumber \end{eqnarray}
\noindent for each possible $\boldsymbol{X_i}, Y_i(0)$ and $Y_i(1)$, where $\phi$ is a vector of parameters controlling this distribution.

Finally, a super-population assignment mechanism is unconfounded if \begin{eqnarray} f_{\boldsymbol{T} | \boldsymbol{Y(0)}, \boldsymbol{Y(1)}, \boldsymbol{X}} (t | y_0, y_1, \boldsymbol{x}, \phi) = f_{\boldsymbol{T} | \boldsymbol{X}}(t|\boldsymbol{x}, \phi)  \ \forall \ y_o, y_1, \boldsymbol{x}, \phi \text{ and } t \in \left \{0, 1\right\}. \nonumber  \end{eqnarray} \noindent For notational convenience, we will drop $\phi$ throughout.

Under an individualistic assignment mechanism, the combination of a probabilistic and unconfounded treatment assignment has been referred to both as strong unconfoundedness and strong ignorability \citep{stuart2010matching}. The class of assignment mechanisms that are individualistic, probabilistic, and unconfounded, but whose control does not lie in the hands of an investigator, are referred to as regular assignment mechanisms, and are most commonly identified with observational data. Weaker versions of unconfoundedness are sufficient for some estimation techniques and estimands \citep{imbens1999role}, and are discussed in Section \ref{matchmultiple}. 

Let $e_{t_1,t_2}(\boldsymbol{X}) = P(T=t_1|\boldsymbol{X})$ be the propensity score (PS), and let $\hat{e}_{t_1, t_2}(\boldsymbol{X})$ be the estimated PS, traditionally calculated using logistic or probit regression. If treatment assignment is regular, then it is possible to estimate unbiased unit-level causal effects  between those at different treatment assignments with equal PS's \citep{rosenbaum1983central}. Propensity scores are often used for either matching, inverse probability weighting or subclassification to estimate (\ref{ate}) and (\ref{att}).

\subsubsection{Description of estimands}

\begin{figure}  
\begin{center} 
\begin{tikzpicture}
    \draw \fourthcircle ;
    \draw \fifthcircle ;
    \begin{scope}[fill opacity=0.5]
      \clip \fourthcircle;
      \fill[gray] \fifthcircle;
    \end{scope}[fill opacity=0.5]
        \draw \fourthcircle ;
        \draw \fifthcircle ;
\node at (0,2.2) {$t_2$};
\node at (-.1,1.6) {$t_1$};
\node at (3,0) { };
\node at (-3,0) { };
\node at (0.2,-.5) {Scenario a};
\end{tikzpicture}
\begin{tikzpicture}
    \begin{scope}[fill opacity=0.5]
    \fill[gray] \thirdcircle;
    \draw \secondcircle ;
    \draw \thirdcircle ;
      \clip \secondcircle;
       \thirdcircle;
    \end{scope}
\node at (0,2.1) {$t_1$};
\node at (1,2.1) {$t_2$};
\node at (3,0) { };
\node at (-1,0) { };
\node at (0.2,-.5) {Scenario b};
        \draw [dashed] \secondcircle ;
        \draw [dashed] \thirdcircle ; 
\end{tikzpicture}
\begin{tikzpicture}
    \begin{scope}[fill opacity=0.5]
    \draw \secondcircle ;
    \draw \thirdcircle ;
      \clip \secondcircle;
      \fill[gray] \thirdcircle;
    \end{scope}
\node at (0,2.1) {$t_1$};
\node at (1,2.1) {$t_2$};
\node at (3,0) { };
\node at (-1,0) { };
\node at (0.2,-.5) {Scenario c};
        \draw [dashed] \secondcircle ;
        \draw [dashed] \thirdcircle ; 
\end{tikzpicture}
\caption{Three scenarios of covariate overlap for binary treatment: shaded areas represent subjects included in a matched analysis}
\label{f21}
\end{center}
\end{figure}
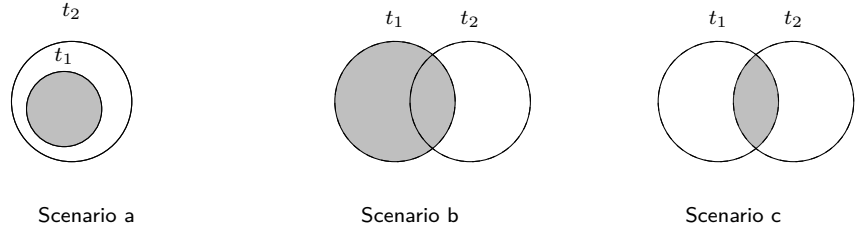

\indent It is useful to describe how estimands are affected by the distribution of $\boldsymbol{X}$ in treatment groups $t_1$ and $t_2$. Figure \ref{f21} shows different sets of overlap in the covariates' distributions between those receiving $t_1$ and $t_2$. Each circle in Figure \ref{f21} represents a hypothetical distribution of $\boldsymbol{X}$ among those exposed to each treatment, allowing for an infinitesimally small number of units outside of it. For example, each circle could represent the 99th percentiles of a two-dimensional multivariate normal distribution. In Figure \ref{f21}, shaded regions correspond to the distribution of covariates in a population of interest, $S_{t_1}$. When $0 \textless P(T=t_2 | X=x*) \textless 1$ $\forall \ x* \in S_{t_1}$, $PATT_{t_1, t_2}$ reflects the $ATT$ of those treated on $t_1$ (Figure \ref{f21}, Scenario a). 

In Scenario b of Figure \ref{f21}, $PATT_{t_1, t_2}$ also intends to reflect the $ATT$ of those receiving $t_1$. However, there exists an $x* \in S_{t_1}$ such that $P(T=t_2 | X=x*) \approx 0$. Thus, the assignment mechanism is not regular and it is impossible to approximate $PATT_{t_1,t_2}$ without making unassailable assumptions due to individuals with covariates lying outside the intersection of the two treatment groups.

One advice to handle this issue is to use a common support region, where those with either $\boldsymbol{X}$ or $\hat{e}_{(t_1, t_2)}(\boldsymbol{X})$ beyond the range of $\boldsymbol{X}$ or $\hat{e}_{(t_1, t_2)}(\boldsymbol{X})$ of those receiving the other treatment are excluded from the analysis phase \citep{dehejia1998causal, crump2009dealing}. A different advice to reduce differences between matched subjects is by using a caliper matching procedure, and dropping units without eligible matches with similar $\hat{e}_{t_1, t_2}(\boldsymbol{X})$ in the other group \citep{caliendo2008some, stuart2010matching}. With either of these advices, the treatment effect only generalizes to those receiving $t_1$ who were eligible to be treated with treatment $t_2$ (i.e., the intersection of the treatment groups in Figure \ref{f21}, Scenario c). Let $E_{1i}$ be an indicator for subject $i$ having a propensity score within the common support of $\hat{e}_{(t_1, t_2)}(\boldsymbol{X})$. Defensible estimands of interest are now
\begin{eqnarray}PATE_{E_1(t_1,t_2)} &=& E [  Y_i(t_1) -  Y_i(t_2) | E_{1i} = 1 ] \label{15}\\
PATT_{E_1(t_1,t_2)} &=& E [  Y_i(t_1) -  Y_i(t_2) | E_{1i} = 1, T_i=t_1 ].\label{16} \end{eqnarray}

\noindent Although estimands (\ref{15}) and (\ref{16}) share the same common support, they may differ if the covariates' distributions of the treated in $E_{1i}$ differ from that of the control.



\subsection{Notation for multiple treatments}

The choice of estimands grows with increasing treatment options. Let $\mathcal{T} = \left \{t_1, t_2, ..., t_Z\right \}$ be the treatment support for $Z$ total treatments, with $\boldsymbol{\mathcal{Y}_i}= \left \{Y_i(t_1), Y_i(t_2), ..., Y_i(t_Z)\right \}$ the set of potential outcomes for subject $i$.

To define potential outcomes and estimate treatment effects with multiple treatments, our assumptions are expanded as follows. First, the SUTVA expands across a subject's vector of potential outcomes. Second, a regular treatment assignment mechanism requires that individualistic, probabilistic, and unconfoundedness hold for multiple exposures.  Like in the binary case, a random sample of $N$ units from an infinite super-population results in an individualistic assignment mechanism. Assignment mechanisms are super-population probabilistic if \begin{eqnarray} 
0 \ \textless \ f_{\boldsymbol{T} | \boldsymbol{Y(t_1)} \ldots \boldsymbol{Y(t_Z)}, \boldsymbol{X}} (t | Y_i(t_1), \ldots, Y_i(t_Z), \boldsymbol{X_i}, \phi) \ \textless \ 1 \ \ \forall \ t \in \left\{t_1, ... t_Z\right\}, \nonumber \end{eqnarray}
for each possible $\boldsymbol{X_i}, Y_i(t_1),\ldots, Y_i(t_Z)$. With multiple treatments, a super-population unconfounded assignment mechanism requires that  \begin{eqnarray} f_{\boldsymbol{T} | \boldsymbol{Y(t_1)}, \ldots , \boldsymbol{Y(t_Z)}, \boldsymbol{X}} (t | y_{t_1}, \ldots y_{t_Z}, \boldsymbol{x}, \phi) = f_{\boldsymbol{T} | \boldsymbol{X}}(t|\boldsymbol{x}, \phi)  \ \forall \ y_{t_1}, \ldots, y_{t_Z}, \boldsymbol{x}, \phi \text{ and } t \in \left \{t_1, \ldots, t_Z \right\}. \nonumber \end{eqnarray}

\noindent We first present a broad definition of the possible contrasts that may be of interest with multiple treatments. Define $w_1$ and $w_2$ as two subgroups of treatments such that $w_1, w_2 \subseteq \mathcal{T}$ and $w_1 \cap w_2 = \emptyset$. Next, let $\norm{w_1}$ and $\norm{w_2}$ be the cardinality of $w_1$ and $w_2$, respectively. Possible estimands of interest are $PATE_{w_1, w_2}$ and $PATT_{w_1|w_1, w_2}$, where
\begin{eqnarray} PATE_{w_1, w_2} &=&  E\bigg[ \frac{\sum_{t\in w_1} Y_i(t)}{\norm{w_1}} - \frac{\sum_{t\in w_2} Y_i(t)}{\norm{w_2}} \bigg] \label{PATEM}, \\
PATT_{w_1|w_1, w_2} &=&  E\bigg[ \frac{\sum_{t\in w_1} Y_i(t)}{\norm{w_1}} - \frac{\sum_{t\in w_2} Y_i(t)}{\norm{w_2}}\ \vrule \ T_i \in w_1  \bigg]. \label{PATTM} \end{eqnarray}

\noindent In (\ref{PATEM}) and (\ref{PATTM}), the expectation is over all units, $i = 1, ... , \mathcal{N}$, and the summation is over the potential outcomes of a specific unit. 

An example of when (\ref{PATEM}) and (\ref{PATTM}) are scientifically meaningful is in a setting with two conventional and three atypical antipsychotic drugs, where physicians first choose drug type (conventional or atypical) before choosing an exact prescription \citep{tchernis2005use}. In this case, an investigator could be interested in the general treatment effect between conventional treatments, $w_1 = \left\{t_1,t_2\right\}$, and atypical ones, $w_2 = \left\{t_3,t_4,t_5\right\}$, and an estimand of interest could be $PATE_{w_1, w_2} = E[\frac{Y_i(t_1)+Y_i(t_2)}{2} - \frac{Y_i(t_3)+Y_i(t_4)+Y_i(t_5)}{3}]$. 

The most traditional estimands with multiple treatments contrast all treatments using simultaneous pairwise comparisons, where $w_1$ and $w_2$ are each composed of one treatment. Using equation (\ref{PATEM}), there are ${Z}\choose{2}$ possible $PATE$'s of interest. It is important to note that pairwise $PATE$'s are transitive. Formally, for $w_1 = \left\{t_1\right\}$, $w_2 = \left\{t_2\right\}$, and $w_3 = \left\{t_3\right\}$, $PATE_{w_1,w_3} - PATE_{w_1,w_2} = PATE_{w_2,w_3}$.

For reference group $w_1 = \left\{t_1\right\}$, researchers are commonly interested in $Z-1$ pairwise $PATT$'s, one for each of the treatments which the reference group did not receive \citep{mccaffrey2013tutorial}. In order to compare among the $Z-1$ treatments, the $PATT$'s should also be transitive, such that $PATT_{w_1|w_1,w_3} - PATT_{w_1|w_1, w_2} = PATT_{w_1|w_2,w_3}$. This property generally does not extend when conditioning on a population eligible for different treatment groups. For example, unless the super populations of those receiving treatments $w_1$ and $w_2$ are identical, $PATT_{w_1|w_1,w_2} - PATT_{w_2|w_2,w_3}$ is generally not equal to $PATT_{w_1|w_1,w_3}$. 

For the remainder of the manuscript, we assume that pairwise contrasts between treatments are the estimands of interest, so that $\norm{w_1} = \norm{w_2} = ... = \norm{w_z} = 1$. 

\subsection{The generalized propensity score} \label{Sec13}

The generalized propensity score (GPS), $r(t,\boldsymbol{X}) = Pr(T=t|\boldsymbol{X=x})$, extends the PS from a binary treatment setting to the multiple treatment setting \citep{imbens1999role, imai2004causal}. 

With a binary treatment, knowing $e_{t_1,t_2}(\boldsymbol{X})$ is equivalent to knowing $1-e_{t_1,t_2}(\boldsymbol{X})$. Thus, two individuals with the same PS are also identical with respect to their probability of receiving $t_2$. Conditioning with multiple treatments, however, often must be done on a vector of GPS's, defined as $\boldsymbol{R(X)} = (r(t_1,\boldsymbol{X}),...,r(t_Z,\boldsymbol{X}))$, or a function of $\boldsymbol{R(X)}$ \citep{imai2004causal}. 

Two individuals with the same $r(t,\boldsymbol{X})$ for treatment $t$ may have differing $\boldsymbol{R(X)}$'s. For example, for $\mathcal{T} = \left\{t_1,t_2,t_3\right\}$, let $\boldsymbol{R(X_i)}, \boldsymbol{R(X_j)},$ and $\boldsymbol{R(X_k)}$ be the GPS vectors for subjects $i$, $j$, and $k$, respectively, where $T_i = t_1$, $T_j = t_2$, and $T_k = t_3$, with 
\begin{eqnarray}
 \boldsymbol{R(X_i)} = (0.30, 0.60, 0.10),\nonumber \\
 \boldsymbol{R(X_j)} = (0.30, 0.35, 0.35),\nonumber \\ 
 \boldsymbol{R(X_k)} = (0.30, 0.10, 0.60).\nonumber  \end{eqnarray}

Even though $r(t_1,\boldsymbol{X_i}) = r(t_1,\boldsymbol{X_j}) = r(t_1,\boldsymbol{X_k}) = 0.30$, because $r(t_2,\boldsymbol{X_i}) \neq r(t_2,\boldsymbol{X_j}) \neq r(t_2,\boldsymbol{X_k})$  and $r(t_3,\boldsymbol{X_i}) \neq  r(t_3,\boldsymbol{X_j}) \neq  r(t_3,\boldsymbol{X_k})$, differences in outcomes between these subjects would generally not provide unbiased causal effect estimates \citep{imbens1999role}. In part due to this limitation, \citet{imbens1999role} called individual matching less 'well-suited' to multiple treatment settings. Only under the scenario of $\boldsymbol{R(X_i)} = \boldsymbol{R(X_j)} = \boldsymbol{R(X_k)}$ would contrasts in the outcomes of subjects $i$, $j$, and $k$ provide unbiased unit-level estimates of the causal effects between all three treatments \citep{imbens1999role, imai2004causal}. 

For nominal treatment, the multinomial logistic and the multinomial probit models have been proposed to estimate $\boldsymbol{R(X)}$, and for ordinal treatment, the proportional odds model has been suggested \citep{imbens1999role, imai2004causal}. Alternatively, researchers have also used models designed for binary outcomes to estimate $\boldsymbol{R(X)}$, including logistic and probit regression models on different subsets of subjects receiving each pair of treatments. Although a multinomial model is more intuitive, in practice, \citet{lechner2002program} identified correlation coefficients of roughly 0.99 when comparing the conditional treatment assignment probabilities from a set of binary probit models to those from a multinomial probit. As another option, \citet{mccaffrey2013tutorial} used generalized boosted models to independently estimate $P(I_i(t)|\boldsymbol{X})$, where $I_i(t) = \left \{1 \text{ if } T_i = t, \ 0 \text{ otherwise}\right \}$. The probabilities estimated using generalized boosted models may not add up to unity. To address this issue, \citet{mccaffrey2013tutorial} proposed an additional procedure that selects one treatment as a holdout and estimates $P(I_i(t) | \boldsymbol{X})$ using the estimated odds ratios of the probability of being assigned to each treatment versus the probability of being assigned to the holdout treatment. The choice of the holdout treatment may result in different estimated probabilities, and because it relies on binary estimation of subsamples of the population, it may not be able to adjust for the entire $\boldsymbol{R(X)}$.  In our review below, we specify the model that is used to estimate the treatment assignment probabilities suggested by each method.



\subsection{Ordinal treatments} \label{SectOrdinal}

With ordinal treatments, such as scales (e.g. never - sometimes - always) or doses (e.g. low - medium - high), it is sometimes possible to condition on a scalar balancing score in place of conditioning on a vector.  This can be done by estimating the assignment mechanism as a function of $\boldsymbol{X}$ using the proportional odds model \citep{mccullagh1980regression}, such that
		\begin{eqnarray}   \text{log}\bigg(\frac{P(T_i \textless t)}{P(T_i \geq t)}\bigg) = \theta_t - \boldsymbol{\beta^TX_i}, t = 1,...,Z-1. \label{MCC} 
		\end{eqnarray} 

\noindent Letting $\boldsymbol{\beta^T} = (\beta_1, .. \beta_p)^T$, \citet{joffe1999invited} and \citet{imai2004causal} showed that after using this model for the assignment mechanism,  differences in outcomes between units with different exposure levels but equal $\boldsymbol{\beta^TX}$ scores can provide unbiased unit-level estimates of causal effects at that $\boldsymbol{\beta^TX}$. 

The balancing property of $\boldsymbol{\beta^TX}$ can be used to match or subclassify subjects receiving different levels of an ordinal exposure. \citet{lu2001matching} used non-bipartite matching to form matched sets based on a function of $\boldsymbol{\beta^TX}$ and the relative distance between exposure levels. While this method does not specify an exact causal estimand, it is used for testing the hypothesis of whether or not a dose-response relationship exists between $T$ and $Y$ (see \citet{armstrong2010chief, frank2010latino, snodgrass2011does} to name a few). 

\cite{imai2004causal, zanutto2005using, yanovitzky2005estimating} and \citet{lopez2014estimating} used Equation (\ref{MCC}) to estimate treatment assignment by subclassifying subjects with similar $\boldsymbol{\beta^TX}$ values. After subclassification on $\boldsymbol{\beta^TX}$, the distribution of $\boldsymbol{X}$ across treatments is roughly equivalent for units in the same subclass. Unbiased causal effects can be estimated within each subclass, and aggregated across subclasses using a weighted average to estimate either $PATE$'s or $PATT$'s \citep{zanutto2005using}. \citet{lopez2014estimating} found that combining regression adjustment with subclassification yielded more precise estimates. 

A different strategy for estimating the causal effects of ordinal exposures is to dichotomize the treatment using a pre-specified cutoff and binary propensity score methods \citep{chertow2004renalism, davidson2006association, schneeweiss2007risk}. This procedure may result in a loss of information, as all subjects on one side of the cutoff are treated as having the same exposure level, and could violate the component of SUTVA which requires no hidden treatment. \citet{royston2006dichotomizing} identified a loss of power, residual confounding of the treatment assignment mechanism, and possible bias in estimates as the results of dichotomization. Moreover, dichotomization makes identification of an optimal exposure level impossible. Thus, matching or subclassifications methods which maintain all exposure levels while balancing on $\boldsymbol{\beta^TX}$ are preferred for causal inference with ordinal exposures \citep{imai2004causal}. Inverse probability weighting can also be used to estimate causal effects from ordinal treatments \citep{imbens1999role}. 

\subsection{Nominal treatments} \label{PEC}

Nominal treatments do not follow a specific order. Thus, it is harder to identify a `sensible' function that reduces $\boldsymbol{R(X)}$ to a scalar. Several methods have been proposed to estimate causal effects with multiple treatments from observational data. We provide an overview of these methods and explicate on their assumptions and estimands.

\subsubsection{Series of binomial comparisons}

\indent \citet{lechner2001identification, lechner2002program} estimated $PATT$'s between multiple treatments using a series of binary comparisons ($SBC$). $SBC$ implements binary propensity score methods within each of the ${Z}\choose{2}$ pairwise population subsets. For example, a treatment effect comparing $t_1$ to $t_2$ uses only subjects receiving either $t_1$ or $t_2$, ignoring subjects that received $t_3$. Lechner advocates matching on either $\hat{e}_{(t_1,t_2)}(\boldsymbol{X})$, estimated using logistic or probit regression, or $\hat{r}(t_1,\boldsymbol{X})/(\hat{r}(t_1,\boldsymbol{X})+\hat{r}(t_2,\boldsymbol{X}))$, where $\hat{r}(t_1,\boldsymbol{X})$ and $\hat{r}(t_2,\boldsymbol{X})$ are estimated using a multinomial regression model.

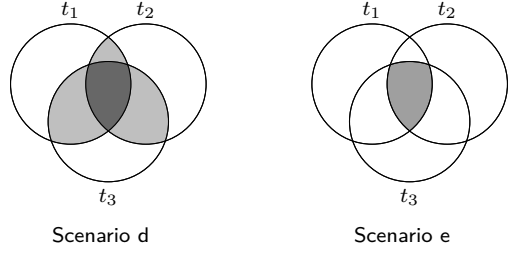
\begin{figure}
\begin{center}
\begin{tikzpicture}
    \draw \firstcircle   ;
    \draw \secondcircle  ;
    \draw \thirdcircle    ;
    \begin{scope}[fill opacity=0.5]
      \clip \firstcircle;
      \fill[gray] \secondcircle;
    \end{scope}
    \begin{scope}[fill opacity=0.5]
      \clip \firstcircle;
      \fill[gray] \thirdcircle;
    \end{scope}
    \begin{scope}[fill opacity=0.5]
      \clip \thirdcircle;
      \fill[gray] \secondcircle;
    \end{scope}
    \begin{scope}[fill opacity=0.5]
      \clip \firstcircle;
      \clip \thirdcircle;
      \fill[darkgray] \secondcircle;
    \end{scope}
        \draw \firstcircle  ;
        \draw \secondcircle  ;
        \draw \thirdcircle  ;
\node at (0,1.99) {$t_1$};
\node at (1,1.99) {$t_2$};
\node at (0.5,-.5) {$t_3$};
\node at (3,2.1) {};
\node at (0.4,-1) {Scenario d};
\end{tikzpicture}
\begin{tikzpicture}
    \draw \firstcircle ;
    \draw \secondcircle   ;
    \draw \thirdcircle   ;
    \begin{scope}[fill opacity=0.5]
      \clip \firstcircle;
      \clip \thirdcircle;
      \fill[darkgray] \secondcircle;
    \end{scope}
        \draw \firstcircle ;
        \draw \secondcircle ;
        \draw \thirdcircle;
\node at (0,1.99) {$t_1$};
\node at (1,1.99) {$t_2$};
\node at (0.5,-.5) {$t_3$};
\node at (0.4,-1) {Scenario e};
\end{tikzpicture}
\end{center}
\caption{Two scenarios of eligible subjects with three treatments: shaded areas represent subjects included in a matched analysis}
\label{f22}
\end{figure}

Figure \ref{f22} (Scenario d) depicts the unique common support regions for $Z=3$ when using SBC, where treatment effects reflect different subsets of the population. Let $\boldsymbol{e_{(t_1,t_2)}(X,T=t_1)}$ and $\boldsymbol{e_{(t_1,t_2)}(X,T=t_2)}$ be the vector of all binary propensity scores among subjects receiving $t_1$ and $t_2$, respectively. We define $E_{2i}(t_1,t_2)$ as the indicator for subject $i$ having a binary propensity score for treatments $t_1$ and $t_2$ within the common support: 
\begin{eqnarray} E_{2i}(t_1,t_2) = \twopartdef { 1 } {e_{(t_1,t_2)}(\boldsymbol{X_i}) \in \boldsymbol{e_{(t_1,t_2)}(X,T=t_1)}\cap  \boldsymbol{e_{(t_1,t_2)}(X,T=t_2)}} {0} {\text{otherwise}} \nonumber\end{eqnarray}

$SBC$ estimates the causal effect of treatment $t_1$ versus treatment $t_2$, among those on $t_1$, as 
\begin{eqnarray} PATT_{E_2(t_1|t_1,t_2)} = E[Y_i(t_1)-Y_i(t_2)|T_i=t_1, E_{2i}(t_1,t_2)=1]. \end{eqnarray} 

Each pairwise treatment effect from $SBC$ generalizes only to subjects eligible for that specific pair of treatments, as opposed to those eligible for all treatments. Such pairwise treatment effects are not transitive, and cannot generally inform which treatment is optimal when applied to the entire population. For example, $PATT_{E_2(t_1|t_1,t_2)}$ and $PATT_{E_2(t_1|t_1,t_3)}$ may generalize to separate subsets of units who received $t_1$ (i.e., the super population where $E_{2i}(t_1|t_1,t_2)=1$ could differ from the super population where $E_{2i}(t_1|t_1,t_3)$=1). 

Despite this major limitation, versions of $SBC$ have been applied in economics, politics, and public health \citep{bryson2002use, dorsett2006new, levin2009measuring, drichoutis2005nutrition, kosteas2010effect}.

\subsubsection{Common referent matching}

\indent With three treatments, \citet{rassen2011simultaneously} proposed common referent matching ($CRM$) to create sets with one individual from each treatment type. For $\mathcal{T} = \left \{t_1,t_2,t_3 \right \}$, the treatment $t_1$ such that $n_{t_1} = \text{min}\left\{n_{t_1},n_{t_2},n_{t_3}\right\}$, is used as the reference group.

$CRM$ is composed of 3 steps. (1) Among those receiving each pair of treatments, $\left \{t_1,t_2\right \}$ or $\left \{t_1,t_3\right \}$, logistic or probit regression is used to estimate $e_{t_1,t_2}(\boldsymbol{X})$ and $e_{t_1,t_3}(\boldsymbol{X})$, respectively; (2) Using 1:1 matching, pairs of units receiving $t_1$ or $t_2$ are matched using $\hat{e}_{t_1,t_2}(\boldsymbol{X})$ and pairs of units receiving $t_1$ or $t_3$ are matched using $\hat{e}_{t_1,t_3}(\boldsymbol{X})$; (3) These two cohorts are used to construct 1:1:1 matched triplets using the patients receiving $t_1$ who were matched to both a unit receiving $t_2$ and a unit receiving $t_3$, along with their associated matches. Matched pairs from treatments $t_1$ and $t_3$ are discarded if the unit receiving $t_1$ was not matched with a unit on treatment $t_2$, and pairs of units receiving $t_1$ and $t_2$ are discarded when there is no match for the reference unit to a unit receiving $t_3$.

Let $E_{3i}$ be the indicator for having two pairwise binary PS's within their respective common supports, such that
\begin{eqnarray} E_{3i} = \twopartdef { 1 } {E_{2i}(t_1,t_2) =1 \text{ and } E_{2i}(t_1,t_3) = 1} {0} {\text{otherwise.}} \nonumber \end{eqnarray}

$CRM$ attempts to estimate the following treatment effects: 
\begin{eqnarray}
	PATT_{E_3(t_1|t_1,t_2)} &=& E[Y_i(t_1)-Y_i(t_2)|T_i=t_1, E_{3i}=1] \nonumber \\ 
	PATT_{E_3(t_1|t_1,t_3)} &=& E[Y_i(t_1)-Y_i(t_3)|T_i=t_1, E_{3i}=1] \nonumber  \\
	PATT_{E_3(t_1|t_2,t_3)} &=& E[Y_i(t_2)-Y_i(t_3)|T_i=t_1, E_{3i}=1] \nonumber \end{eqnarray}

\noindent  $PATT_{E_3(t_1|t_2,t_3)}$ is the average difference in the potential outcomes of receiving treatments $t_2$ and $t_3$ among the population of subjects who received $t_1$. 

\citet{rassen2011simultaneously} relied on common sampling variance estimates produced by the SAS statistical software \citep{SASSTAT} to make inference. These estimates may underestimate the sampling variance, because they ignore the variability induced by the matching procedure. The next section will explain the possible issues that arise from $CRM$ and similar procedures.

\subsubsection{Interlude: Binary PS applications to multiple treatments} \label{BinaryPS}

The following hypothetical example with $Z=3$ illustrates issues with the implementation of binary PS tools, as in $SBC$ and $CRM$, when there are multiple treatments.

Let $X_i = {x_{1i} \choose x_{2i}} $ be a vector of covariates for subject $i$, and we will assume that $X_i|T_i =t \sim N(\mu_t, \mathbb{1})$, where $\mu_t$ is a 2 x 1 mean vector and $\mathbb{1}$ is the 2 x 2 identity matrix. Further, we let $\mu_1 = (0, 0)$, $\mu_2 = (0, a)$, and $\mu_3 = (a, 0)$. 

An arbitrary linear combination of $X$ can be expressed as the sum of components along the standardized linear discriminant, $\mathcal{Z}$, and orthogonal to it, $\mathcal{W}$ \citep{rubin1992affinely}. Matching on the true or estimated propensity score does not introduce any bias in $\mathcal{W}$ when $X_i |T_i$ follows a multivariate normal distribution. In addition, after matching, $\mathcal{W}$ will have the same expected second moment \citep{rubin1992characterizing}. Specifically, when matching treatment 1 to treatment 2 with $a=2$, $Z_{12} = {0 \choose 2} ' X_1 / \sqrt{2} = \sqrt{2} X_2$ and $\mathcal{W}_{12} = X_1$. After matching, \cite{rubin1992characterizing} showed that 

\begin{eqnarray}
 E(\overline{\mathcal{Z}}_{12}^{m_2}) &=& 2 - \Omega (N_{t_2}, n_{t_2}) \cong 2 - 2\pi \text{log}(\frac{N_{t_2}}{n_{t_2}}) \nonumber \\
 E(\overline{\mathcal{Z}}_{12}^{m_1}) &=& 0+ \Omega (N_{t_1}, n_{t_1}) \cong 2 + 2\pi \text{log}(\frac{N_{t_1}}{n_{t_1}}) \nonumber \end{eqnarray}

\noindent where $\overline{\mathcal{Z}}_{12}^{m_1}$ and $\overline{\mathcal{Z}}_{12}^{m_2}$ are the averages of the standardized linear discriminate in the matched treatments 1 and 2, respectively,  $\Omega (N_{t}, n_{t})$ is the average expectation of the $n$ largest of the $N$ randomly sampled standard normal variables, and its approximation was depicted in \cite{rubin1976multivariate}. 

In our example with $a=2$, $\mu_m =  E(\overline{\mathcal{Z}}_{12}^{m_1}) =  E(\overline{\mathcal{Z}}_{12}^{m_2}) $ when $\frac{N_{t_2}}{n_{t_2}}$ and $\frac{N_{t_1}}{n_{t_1}}$ are bigger than 3. Similar results can be derived when matching treatments 1 and 3 with $Z_{13} = \sqrt{2} X_1$ and $\mathcal{W}_{13} = X_2$.

Matching units that received either treatment 1 or 2 separate from units that received either treatment 1 or 3 generates two subpopulations, one with mean ${0 \choose \mu_m}$ and another with mean ${\mu_m \choose 0}$.  Note that $\overline{\mathcal{W}}_{12}^{m_1}$ and $\overline{\mathcal{W}}_{12}^{m_2}$ are independent and have similar means \citep{rubin1992characterizing}. Similarly, $\overline{\mathcal{W}}_{13}^{m_1}$ and $\overline{\mathcal{W}}_{13}^{m_3}$ are independent and have similar means. Lastly, $\overline{\mathcal{W}}_{12}^{m_1}$ is independent from $\overline{\mathcal{W}}_{13}^{m_1}$. When using $CRM$, the units that are kept as matches that received treatment 1 will have the high values of $X_1$ and $X_2$. However, because of the independence, group 2 will still have $\overline{\mathcal{W}}_{12}^{m_2}$ that has a mean close to zero and group 3 will still have $\overline{\mathcal{W}}_{13}^{m_3}$ that has a mean close to zero. Thus, in certain settings $CRM$ may perform worse than without matching. 

This analysis can be observed in a simple simulation where, letting $a$ = 2, $n_{t_1}$= 400, and $n_{t_2} = n_{t_3} = 800$, we calculate the sample means among those matched after using a binary matching algorithm (with caliper 0.25*SD($e_{t_1,t_2}(\boldsymbol{X})$)). Table \ref{TabSimEarly} shows the median covariate values among those receiving each treatment, using only the subjects that remain after matching. 

\begin{table} [h]\centering \small 
\caption {Median covariate values among those matched using a binary algorithm with $Z=3$}
\label{TabSimEarly}
\begin{tabular}{l r r}
\hline 
  $T$  & $X_1$  &$X_2$ \\ \hline
$t_1$ & 0.71 (0.56, 0.82) & 0.72 (0.57, 0.83) \\
$t_2$ & 0.70 (0.56, 0.84) &0.01 (-0.16, 0.20) \\
$t_3$ & 0.01 (-0.19, 0.18)& 0.72 (0.58, 0.85) \\ \hline 
\multicolumn{3}{l}{2.5th, 97.5th percentiles shown in parenthesis} \\ \hline
\end{tabular}
\end{table}
\noindent Among the matched set, those receiving $t_1$ are similar to those receiving $t_2$ on $X_1$ but not $X_2$, and similar to those receiving $t_3$ on $X_2$ but not $X_1$. 

\begin{figure}[htbp]
\begin{center}
  \includegraphics[width=5.81in,height=4.2in,keepaspectratio]{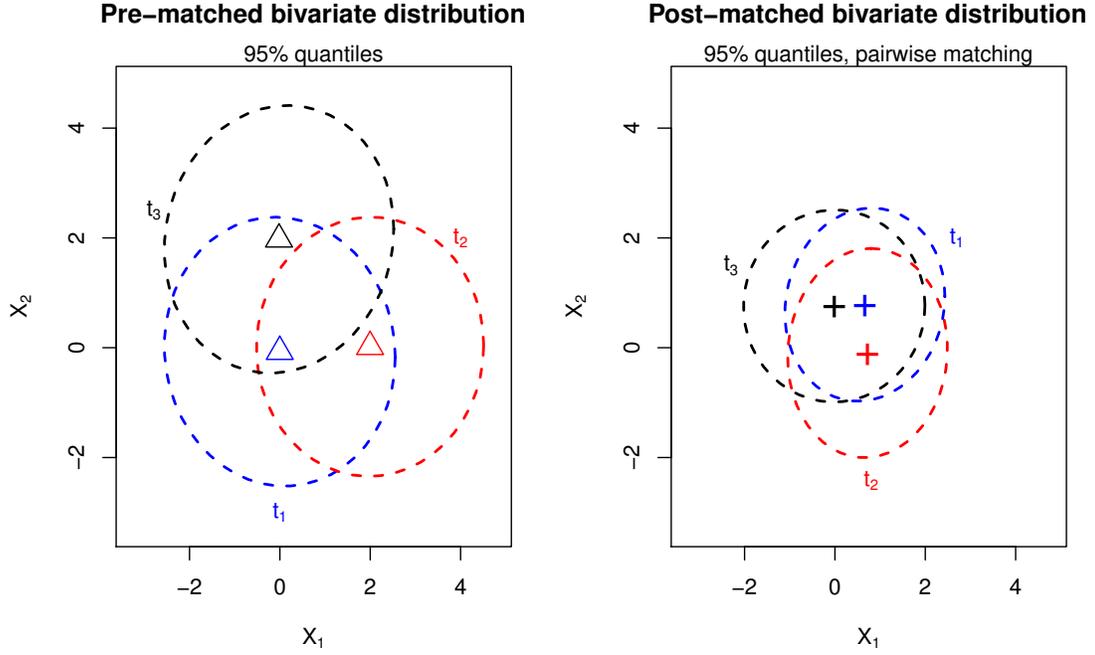}
\caption{95\% quantiles of bivariate $X_1$ and $X_2$ distribution among subjects matched for $Z=3$, for pre-matched (left) and post-matched (right) cohorts. Means of the pre and post-matched covariates' distributions depicted by symbols.}
  \label{FigSimEarly}
\end{center}
\end{figure}

Figure \ref{FigSimEarly} depicts one iteration. The ellipses represent 95\% quantiles of the bivariate distribution of $X_1$ and $X_2$, with one ellipse for subjects receiving each treatment both before and after matching. The triangles represent the pre-matched sample mean among those receiving each treatment, while the `+` signs are the mean covariate values among those matched. While matching reduced the covariates' bias relative to the pre-matched sample, the covariate spaces of those receiving each treatment remain unique in the post-matched cohort, and there is limited overlap between subjects receiving $t_2$ and $t_3$.

\subsubsection{Inverse probability weighting for multiple treatments} \label{matchmultiple}

One common approach for estimating causal effects with multiple treatments uses the inverse probability of treatment assignment as weights \citep{imbens1999role, feng2012generalized, mccaffrey2013tutorial}. When estimating the $PATE$ and $PATT$ with IPW, a relaxed version of the assumption of a regular treatment assignment can be adopted. IPW requires only that $\forall \text{ } t \in \mathcal{T}$, $P(I_i(t)=1|Y_i(t),X_i) = P(I_i(t)=1|X_i)$ to estimate $PATE_{t_1, t_2}$ and $PATT_{t_1, t_2}$. This condition is referred to as weak unconfoundedness instead of strong unconfoundedness \citep{imbens1999role}. \citet{imbens1999role} acknowledges that the contrast between weak unconfoundedness and strong unconfoundedness is `not very different.'

\citet{feng2012generalized} implemented $IPW$ to estimate $PATE$'s between each pair of treatments, such that to contrast $t_1$, $t_2 \in \mathcal{T} $, 
\begin{eqnarray}\widehat{PATE_{t_1,t_2}} &=&\widehat{E[Y_i(t_1)]} -\widehat{E[Y_i(t_2)]} \text{ where}  \label{Feng} \\
 \widehat{E[Y_i(t_1)]} &=& \Bigg( \sum_{i=1}^N \frac{I(T_i=t_1) Y_i}{r(t_1,\boldsymbol{X_i})}\Bigg) \Bigg(\sum_{i=1}^N \frac{I(T_i=t_1)}{r(t_1,\boldsymbol{X_i})}\Bigg)^{-1} \text{ and} \nonumber  \\
 \widehat{E[Y_i(t_2)]} &=& \Bigg( \sum_{i=1}^N \frac{I(T_i=t_2) Y_i}{r(t_2,\boldsymbol{X_i})}\Bigg) \Bigg(\sum_{i=1}^N \frac{I(T_i=t_2)}{r(t_2,\boldsymbol{X_i})}\Bigg)^{-1}. \nonumber \end{eqnarray}

When using $IPW$, extreme weights that are close to 0 can yield erratic causal estimates with large sample variances \citep{little1988missing, kang2007demystifying, stuart2008best}, an issue which is increasingly likely as $Z$ increases, where treatment assignment probabilities for some treatments may become quite small. For example, in an analysis of rare treatment decisions with $Z=7$, \citet{kilpatrick2012exploring} found weights greater than $10^4$ and resulting confidence intervals that were sensitive to model specification. A possible solution to the unstable estimates that has been applied in the binary treatment setting is to trim subjects with extreme weights \citep{lee2011weight}. \citet{kilpatrick2012exploring} observed increased precision with weight removal, relative to the inclusion of all subjects; however, dropping extreme weights also yielded increased bias. This observation reveals a subtle point that is not always recognized. As shown in Section \ref{Sec13}, in contrast to binary propensity scores, the comparison of units with similar $r(t, \boldsymbol{X})$ and different $\boldsymbol{R(X)}$ that receive different treatments has no causal interpretation \citep{imbens1999role}. Instead, only a comparison of the $r(t, \boldsymbol{X})$ weighted averages has such interpretation. As a result, trimming units with $r(t, \boldsymbol{X})$ that are close to 0 or 1 may actually drop units with different covariates' distributions, which could ultimately increase the bias.

For binary treatment, other approaches have been suggested to limit the effects of large weights. These include a doubly robust approach \citep{tan2010bounded}, a covariate balancing propensity score \citep{imai2014covariate}, and generalized boosted models \citep{mccaffrey2004propensity, mccaffrey2013tutorial}, with the latter two methodologies also extending to a multiple treatments framework. To provide confidence intervals for (\ref{Feng}), \citet{feng2012generalized} use the 2.5 and 97.5 quantiles from a non-parametric bootstrap algorithm \citep{efron1994introduction} to obtain a 95\% confidence interval, while \citet{mccaffrey2013tutorial} approximate the standard errors by using robust (or so-called `sandwich') procedure. However, \citet{mccaffrey2013tutorial} acknowledge that there is currently no theory that guarantees that these will result in proper confidence intervals when using generalized boosted models, and this is an area for further statistical research.

\subsubsection{Matching for multiple treatments} \label{matchmultiple2}

Recently, attempts have been made to group several subjects together who have similar $\boldsymbol{R(X)}$, including at least one subject receiving each treatment. With $Z=3$, \citet{rassen2013matching} proposed `within-trio' matching ($WithinTrio$) to form triplets of subjects. $WithinTrio$ uses the KD-tree algorithm \citep{moore1991introductory} to optimize triplet similarities based on units' GPS's for treatments $t_1$ and treatments $t_2$, by using a distance function between all possible pairs of triplets \citep{hott2012division}. Using simulations, \citet{rassen2013matching} found that triplets produced using $WithinTrio$ generally yielded lower standardized covariate bias when compared to $CRM$ and $SBC$. 

One limitation of $WithinTrio$ is that it uses only $t_1$ as the reference treatment, where $n_{t_1} = \text{min}\left\{n_{t_1},n_{t_2},n_{t_3}\right\}$, and so $PATT$'s generalizable to those receiving treatment $t_2$ or $t_3$ cannot yet be estimated. Because all subjects receiving $t_1$ are matched, there is also the potential to form dissimilar triplets, if, for example, all close matches to a subject who received $t_1$ are already taken as matches by other subjects. At this stage in its development, $WithinTrio$ has  focused on $Z = 3$ treatment types. An additional limitation is that there is no known procedure for sampling variance estimates, and application of the bootstrap method may be computationally intensive.


\citet{tu2012comparison} examined a clustering algorithm to bin units into subclasses based on their $\widehat{\boldsymbol{R(X)}}$'s using simulations. The authors showed that $K-means$ clustering ($KMC$, \cite{johnson1992applied}) on the logit transformation of the GPS vector, $\text{logit} \bigg( \widehat{\boldsymbol{R(X)}}\bigg) = \bigg(\text{log}(\widehat{r(t_1,\boldsymbol{X}})/(1-\widehat{r(t_1,\boldsymbol{X})})),...,\text{log}(\widehat{r(t_Z,\boldsymbol{X})}/(1-\widehat{r(t_Z,\boldsymbol{X})})\bigg)$, generally provided the highest within subclass covariate similarity between those receiving different treatments. Although the authors do not provide guidelines regarding which units should be included in generating the clusters (e.g., a common support), if all subjects were subclassified, causal effects could be estimated within each subclass and then aggregated across subclasses using a weighted average to estimate either $PATE$'s or $PATT$'s. One possible issue with clustering on $\boldsymbol{R(X)}$ is that some subclasses may not include units from all treatment groups, which will require extrapolation to that subclass. We know of no implementations of $KMC$ to estimate causal effects for a nominal exposure with real data. Moreover, there is no known procedure for estimating the sampling variance, and randomization based sampling variance estimates may be too small \citep{gutman2015estimation}.

\section{Matching on a vector of generalized propensity scores}  \label{VM}

In observational studies that intend to compare multiple treatments, matching algorithms attempt to eliminate extraneous variation due to observed covariates. In other words, matching attempts to replicate a multi-arm randomized trial where the covariates' distributions of units in each arm are similar. When the number of covariates is significantly larger than the number of treatments, matching on the GPS can reduce the complexity of the algorithms in comparison to matching on the complete set of covariates.

As was shown in Section \ref{BinaryPS}, relying on standard matching tools for two treatments may result in treatment groups with different distributions of covariates, because matching on a single treatment assignment probability does not ensure similarity across the GPS vector. Additionally, approaches like $SBC$ and $CRM$ generalize to specific pairwise subsets of the population, which may be insufficient for clinicians and policy makers, who are generally looking to compare three or more active treatments at once \citep{rassen2011simultaneously, hott2012division}. Meanwhile, current approaches designed to match for multiple treatments tend to be either inaccessible or limited in scope.

To address these limitations, we propose a new algorithm, called vector matching ($VM$), which can match subjects with similar $\boldsymbol{R(X)}$ using available software. $VM$ is designed to generalize to subjects `eligible' for all treatments simultaneously, which is representative of the multi-arm clinical trial that we are hoping to replicate. We begin by describing the treatment effect that we estimate using $VM$.
 
 \subsubsection{Estimands and common support} 

We expand the work of \cite{dehejia1998causal} to identify a common support for multiple treatments as follows. Estimate $\boldsymbol{R(X)}$ using, for example, a multinomial regression model. For each treatment $t\in \mathcal{T}$, let 
\begin{eqnarray}
	r(t,\boldsymbol{X})^{(low)} &=& max\bigg(min(\boldsymbol{r(t,X|T=t_1)}),...,min(\boldsymbol{r(t,X|T=t_Z)})\bigg) \\
	r(t,\boldsymbol{X})^{(high)} &=& min\bigg(max(\boldsymbol{r(t,X|T=t_1)}),...,max(\boldsymbol{r(t,X|T=t_Z)})\bigg) \end{eqnarray}
\noindent where $\boldsymbol{r(t,X|T=\ell)}$ is the treatment assignment probability for $t$ among those who received treatment $\ell$. This is a rectangular common support region that may drop some units that could be included in the analysis. A more complex common support region based on multidimensional ellipsoids or convex hull regions provide areas for further research.
 
Subjects with $r(t,\boldsymbol{X}) \not \in \big(r(t,\boldsymbol{X})^{(low)},(r(t,\boldsymbol{X})^{(high)}\big)$ $\forall \text{ }t \in \mathcal{T}$ may have $\boldsymbol{X}$ values that are not observed for some treatment groups, and should be discarded. After using this exclusion criterion, it is recommended to re-fit the GPS model, to ensure that estimated GPS's are not disproportionately impacted by those dropped (adapted from the binary treatment scenario in \citet{ImbensRubinBook}). Re-fitting is generally done once; unless the minimum and maximum estimated GPSs are identical among each group receiving each treatment, there will always be subjects outside the boundaries in a continuously re-fit model.

Let $E_{4i}$ be the indicator for all treatment eligibility, where
\begin{eqnarray} E_{4i} = \twopartdef { 1 } {r(t,\boldsymbol{X_i}) \in \big(r(t,\boldsymbol{X})^{(low)},(r(t,\boldsymbol{X})^{(high)}\big) \text{  } \forall \text{ } t \in \mathcal{T}} {0} {\text{otherwise}} \nonumber\end{eqnarray}
\noindent The shaded region in Figure \ref{f22}, Scenario e, depicts the subset of those eligible for all three treatments. 

Using $t_1$ as a reference treatment, $PATT$'s among subjects eligible for all treatments are defined as follows. 
\begin{eqnarray} PATT_{E_4(t_1|t_1,t_2)} &=& E[Y_i(t_1)-Y_i(t_2)|T_i=t_1, E_{4i}=1] \label{rs} \\ 
	PATT_{E_4(t_1|t_1,t_3)} &=& E[Y_i(t_1)-Y_i(t_3)|T_i=t_1, E_{4i}=1]  \nonumber  \\
	  ...		&=& ... \nonumber \\    
	PATT_{E_4(t_1|t_1,t_Z)} &=& E[Y_i(t_1)-Y_i(t_Z)|T_i=t_1, E_{4i}=1]  \label{rz}     \end{eqnarray}

There are two benefits to our definition of eligibility. First, all estimands in (\ref{rs}) are transitive; $PATT_{E_4(t_1|t_1,t_2)}$ and $PATT_{E_4(t_1|t_1,t_3)}$, for example, could be contrasted to compare $t_2$ and $t_3$ in the population of subjects who received $t_1$. Second, because all subjects included have $r(t,\boldsymbol{X})^{(low)} < r(t,\boldsymbol{X}) < r(t,\boldsymbol{X})^{(high)} \ \forall \ t$, extrapolation to subjects that did not received a specific treatment is reduced. 

\subsection{Vector Matching} \label{VMSection}

As described in Section \ref{Sec13}, when comparing multiple treatments, the GPS is a vector composed of $Z-1$ independent components; ultimately, our goal is similarity across this vector. One possible matching algorithm for $\boldsymbol{R(X)}$ begins by creating K1 intervals based on $r(t_1, \boldsymbol{X})$ so that there is at least one unit from each treatment group in each interval. The algorithm continues by subclassifying units into K2 intervals within each of the K1 intervals with similar $r(t_2, \boldsymbol{X})$ such that each new interval includes at least one unit from each treatment group. This proceeds until all of the components of $\boldsymbol{R(X)}$ have been subclassified. Such an algorithm may be influenced by the order that the components of $\boldsymbol{R(X)}$ are subclassified. Some orderings of the components may lead to declaring a large set of units as unmatchable and may result in estimates that have limited use in practice.
 
To handle these difficulties, vector matching consists of two steps that can be implemented using common software. First, place subjects into clusters using $KMC$ such that subjects within each cluster are roughly similar on one or more GPS components and there is at least one subject from each treatment in each cluster. Second, match pairs of subjects together only if they appear in the same subclass. 

Below, we explicate and summarize the procedure for a reference treatment $t \in \mathcal{T} = \left\{t_1,...t_Z\right\}$.

\begin{enumerate} 
  \item Estimate $\boldsymbol{R(X_i)}, i = 1, ..., N$ using, for example, a multinomial logistic model.
  \item Drop units outside the common support (e.g., those with $E_{4i} = 0$), and re-fit the model once. 
  \item $\forall \ t' \ne t$ 
	\begin{enumerate} \item Classify all units using $KMC$ on the logit transform of $\boldsymbol{\widehat{R}_{t,t'}(\boldsymbol{X}})$, where $\boldsymbol{\widehat{R}_{t,t'}(\boldsymbol{X}}) = (\widehat{r(\ell,\boldsymbol{X}}) \ \forall \ \ell \neq t,t' )$. This forms $K$ strata of subjects, with similar $Z-2$ GPS scores (not including $\widehat{r(t,\boldsymbol{X}})$ or $\widehat{r(t',\boldsymbol{X}})$) in each $k \in K$. 
     \begin{itemize} \item $Example$: with $Z=5$, $\mathcal{T} = \left\{t_1,..t_5\right\}$, reference treatment $t_1$ and letting $t'=t_2$, $VM$ would use $KMC$ on logit($\widehat{r(t_3,\boldsymbol{X_i})},\widehat{r(t_4,\boldsymbol{X_i})},\widehat{r(t_5,\boldsymbol{X_i})})$ \end{itemize}
  \item Within each strata $k\in K$, use 1:1 matching to match those receiving $t$ to those receiving $t'$ on logit($\widehat{r(t,\boldsymbol{X_i})}$). Matching is performed with replacement using a caliper of $\epsilon*$SD(logit($\widehat{r(t,\boldsymbol{X_i})}$)), where $\epsilon$ = 0.25. 
  \begin{itemize} \item  $Example$: this matches subjects receiving $t_1$ to those receiving $t_2$ within each of the strata produced by $KMC$ \end{itemize}
 \end{enumerate}  \item Subjects receiving $t$ who were matched to subjects receiving all treatments $\ell \neq t$, along with their matches receiving the other treatments, compose the final matched cohort.   
\end{enumerate}
\noindent Up to $n_{t_1,E_4=1}$ sets can be generated using vector matching, where $n_{t_1,E_4=1}$ is the number of subjects receiving $t_1$ with $E_{4i}=1$. 

For $Z=3$, vector matching reduces to: 

\begin{enumerate} 
   \item Match those receiving $t_1$ to those receiving $t_2$ on logit($\widehat{r(t_1,\boldsymbol{X_i})}$) within $K-means$ strata of logit($\boldsymbol{\widehat{r(t_3,\boldsymbol{X_i})}}$) 
   \item Match those receiving $t_1$ to those receiving $t_3$ on logit($\widehat{r(t_1,\boldsymbol{X_i})}$) within $K-means$ strata of logit($\boldsymbol{\widehat{r(t_2,\boldsymbol{X_i})}}$) 
  \item Extract the subjects receiving $t_1$ who were matched to both subjects receiving $t_2$ and $t_3$, as well as their matches.
\end{enumerate}


After the completion of VM, we are left with many sets that include a unit from the reference treatment and matched units from each of the other $Z-1$ treatments. By matching within a subclass, we have ensured that matched units are close on one component of the GPS and roughly similar on the other components. As a result, $VM$ improves the balance in covariates' distributions between those receiving different treatments relative to matching on a single element of the GPS. $VM$ is relatively efficient computationally, and is not as affected by the ordering of the GPS elements.

We implemented $VM$ by matching on logit($\widehat{r(t,\boldsymbol{X})}$) as well as $\widehat{r(t,\boldsymbol{X})}$ within strata estimated using KMC. The logit transformation produced smaller biases, which parallels findings observed with binary treatment \citep{rosenbaum1985constructing}. Additionally, while the recommendation for binary treatment uses $\epsilon = 0.25$ \citep{austin2011optimal}, we examined $\epsilon \in \left\{0.25, 0.50, 1.0\right\}$. Based on the simulation design that is described in Section \ref{Sim}, $VM$ performed best in terms of bias and percent of matched eligible subjects with $\epsilon = 0.25$ (data not shown). The in strata matching procedure is implemented using the $Matching$ \citep{sekhon2008multivariate} package in $R$ statistical software \citep{Rstats}.

\begin{figure}[htbp]
\begin{center}
  \includegraphics[width=5.81in,height=4.4in,keepaspectratio]{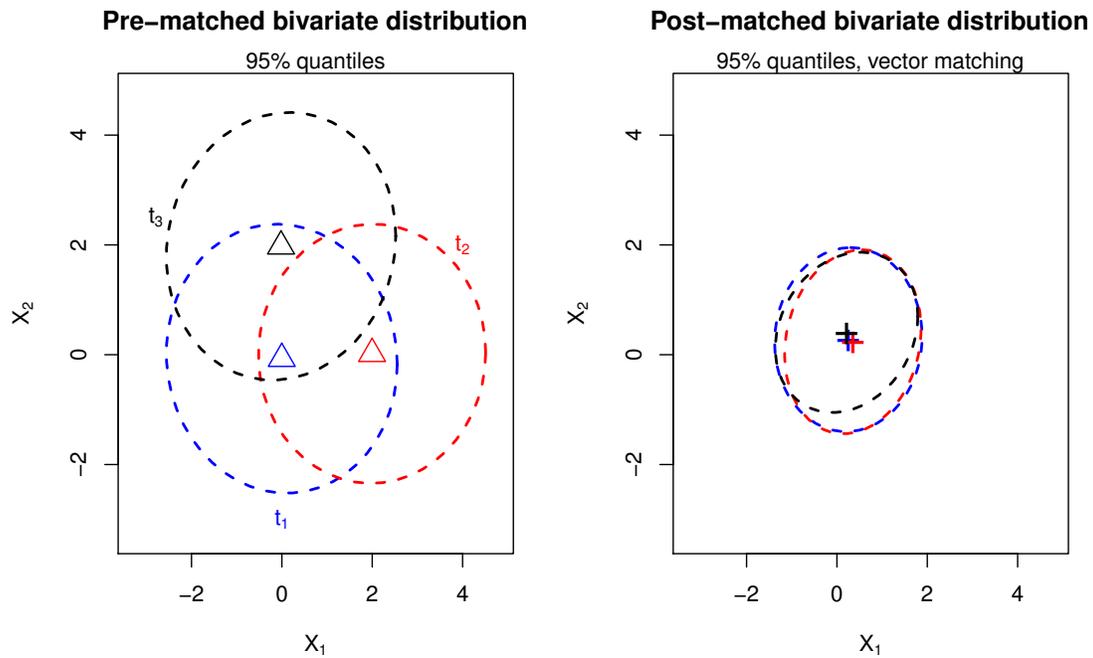}
\caption{95\% quantiles of bivariate $X_1$ and $X_2$ distribution among subjects matched for $Z=3$, for pre-matched (left) and post-matched (right) cohorts. Means of the pre and post-matched distributions depicted by symbols.}
  \label{FigSimLate}
\end{center}
\end{figure}

Figure \ref{FigSimLate} shows the 95\% quantiles of the bivariate $X_1$ and $X_2$ distribution after implementing vector matching on the same iteration as the one shown in Figure \ref{FigSimEarly} (Section \ref{BinaryPS}). Whereas binary procedures were insufficient for identifying similar matched sets, the circles are near perfect overlaps after using vector matching.

\subsection{Post-matching analysis} \label{Analysis}

Although our focus is on the design phase of matching for multiple treatments, it is important to consider how matched sets could be used to make inferences. Point estimates for (\ref{rs}) - (\ref{rz}) using $VM$ matches can be obtained by contrasting those matched using a weighted average, with weights proportional to  $\psi_i$, where $\psi_i$ is the number of times subject $i$ is part of a matched set. Let $n_{trip}$ be the number of matched sets. Point estimates for (\ref{rs}) - (\ref{rz}) can be obtained using (\ref{att_est}) - (\ref{att_est2}), where

\begin{eqnarray}
SATT_{E_4(t_1|t_1,t_2)}  & = &       \frac{ \sum_{i \in E_4}   Y_{i}I(T_i = t_1)\psi_{i} -  Y_{i}I(T_i = t_2)\psi_{i}}{n_{trip}} \label{att_est} \\
	SATT_{E_4(t_1|t_1,t_3)} &=&  \frac{ \sum_{i \in E_4}   Y_{i}I(T_i = t_1)\psi_{i} -  Y_{i}I(T_i = t_3)\psi_{i}}{n_{trip}} \nonumber  \\
	  ...		&=& ... \nonumber \\    
	SATT_{E_4(t_1|t_1,t_Z)} &=&  \frac{ \sum_{i \in E_4}   Y_{i}I(T_i = t_1)\psi_{i} -  Y_{i}I(T_i = t_Z)\psi_{i}}{n_{trip}} \label{att_est2}.    \end{eqnarray}
	
\noindent As highlighted earlier, an advantage of these estimands is that they generalize to subjects `eligible` for all treatments.

Like other approaches that match with multiple treatments, estimating the standard error of these point estimates is still an open research question. One approach for estimating the sampling variances of (\ref{att_est}) - (\ref{att_est2}) is to use functions of the sample variances of $\boldsymbol{Y}$ within those matched at each treatment group. \cite{hill2006interval} provide weighted variance formulas where the variance in each treatment group is weighted to account for multiplicities in the matched units. \cite{abadie2006large} derived a different weighted consistent estimator for the sampling variance of the $PATE$ and the $PATT$ for a binary treatment. Their estimator matches units with similar covariates' values within each treatment group to estimate the variability of the unit level effects. In general, weighted variance estimators may overestimate the true sampling variance, because they do not account for the correlation between subjects that are matched to one another. Deriving closed form solutions for multiple treatments is an area for further research.

Bootstrapping was proposed as a possible technique to estimate the standard errors of matching estimators of the $PATE$ and $PATT$ in a binary treatment setting. For matching without replacement, \cite{austin2014use} identified that a bootstrap algorithm that sampled the matched pairs resulted in estimates of the standard error that were close to the empirical standard deviation of the sampling distribution of the estimated treatment effect. For matching with replacement, \cite{hill2006interval} proposed a more complex form of the bootstrap algorithm. In the complex bootstrap algorithm, bootstrap samples from the original sample are drawn, and within each bootstrap sample, a separate propensity score model is fit and unique sets of matches are identified. In a simulation analysis, the complex bootstrap method was shown to be statistically valid without having extremely large average interval lengths. A similar strategy could be employed with multiple treatments by using $VM$ within each iteration of the bootstrap. However, we caution against use of a similar procedure, because in the binary treatment setting, the bootstrap procedure can either overestimate or underestimate the asymptotic variance given that there can be a high degree of consistency in subjects that are matched to one another after using with-replacement matching \citep{abadie2008failure}.

A different computationally intensive strategy is to use randomization-based approaches, in which the distributions of treatment effects under the null are formed using different permutations of treatment assignments. \cite{rosenbaum2002observational} described such a permutation approach in the context of matching with a binary treatment and non-overlapping sets of matches. \cite{hill2006interval} implemented a similar strategy when matching with replacement by using the Hodges-Lehmann aligned rank test. In simulation analysis, they showed that this test outperformed both the bootstrap and the weighted variance estimators for with replacement matching. Extending this approach for multiple treatments is possible; each matched set obtained by $VM$ would be permuted independently, with the observed test statistic compared to the randomization distribution obtained by these permutations.  For multiple treatments with matched cohorts, the Friedman test statistic \citep{sprent2007applied} or the Quade test statistic \citep{quade1979using} can be used as alternatives to the Hodges-Lehman aligned rank test statistic.
 
\section{Simulations} \label{Sim}

We examine the performance of the methods described in Section \ref{PEC} and the newly proposed method in reducing the bias on observed $\boldsymbol{X}$ using simulations. SBC is not included in the analysis because it cannot be used to contrast three or more treatments simultaneously. Additionally, we assume no natural ordering to the treatment, and thus methods designed for ordinal treatments (Section \ref{SectOrdinal}) are excluded. 

\subsection{Evaluating balance of matched sets by simulation}

In order to provide advice to investigators and following \cite{rubin2001using}, we generated simulation configurations that are either known or can be estimated from the data. A P-dimensional $\boldsymbol{X}$ was generated for $N = n_{t_1} + n_{t_2} +n_{t_3} $ subjects receiving one of three treatments, $\mathcal{T} \in \left\{t_1,t_2,t_3\right\}$, with $n_{t_1}$, $n_{t_2} = \gamma n_{t_1}$, and $n_{t_3} = \gamma^2n_{t_1}$ the sample size of subjects receiving treatments $t_1$,$t_2$,and $t_3$. For a similar set of simulations using $Z$ = 5, see Appendix \ref{AppA}. The values of $\boldsymbol{X}$ were generated from multivariate symmetric distributions such that

\begin{eqnarray} T_i &=& t_1, i = 1, ...,n_{t_1}\\
 	 T_i &=&t_2, i = n_{t_1}+1, ..., n_{t_1}+\gamma n_{t_1} \nonumber \\
	  T_i &=& t_3, i = n_{t_1}+\gamma n_{t_1} + 1, ...,n_{t_1}+\gamma n_{t_1}+\gamma ^2n_{t_1} \nonumber  \end{eqnarray}
\begin{eqnarray}
	\boldsymbol{X_i} |\left\{T_i=t_1\right\}& \sim& f(\boldsymbol{\mu_1}, \boldsymbol{\Upsigma_1}),i = 1, ..., n_{t_1} \\
	\boldsymbol{X_i} |\left\{T_i=t_2\right\} &\sim& f(\boldsymbol{\mu_2}, \boldsymbol{\Upsigma_2}), i = n_{t_1}+1, ..., n_{t_1}+\gamma n_{t_1} \\
	\boldsymbol{X_i} |\left\{T_i=t_3\right\} &\sim& f(\boldsymbol{\mu_3}, \boldsymbol{\Upsigma_3}), i = n_{t_1}+\gamma n_{t_1} + 1, ..., n_{t_1}+\gamma n_{t_1}+\gamma ^2 n_{t_1} \end{eqnarray}
\begin{eqnarray}	\mu_1= ((b,0,0),...,(b,0,0))^T \label{mu1},	\mu_2 = ((0,b,0),...,(0,b,0))^T \label{mu2} \text{, and }	\mu_3 =((0,0,b),...,(0,0,b))^T \label{mu3} \end{eqnarray}
\begin{eqnarray}	\Upsigma_1 = \left( \begin{array}{cccc}  
		1&\tau&...&\tau\\
		\tau&1&...&\tau\\
		.&.&...&.\\
		\tau&\tau&...&1\\ \end{array} \right), 
\Upsigma_2 =\left( \begin{array}{cccc}  
		\sigma_2&\tau&...&\tau\\
		\tau&\sigma_2&...&\tau\\
		.&.&...&.\\
		\tau&\tau&...&\sigma_2\\ \end{array} \right), \text{and }
\Upsigma_3 = \left( \begin{array}{cccc}  
		\sigma_3&\tau&...&\tau\\
		\tau&\sigma_3&...&\tau\\
		.&.&...&.\\
		\tau&\tau&...&\sigma_3\\ \end{array} \right) \end{eqnarray}
The following design implicitly assumes a regular assignment mechanism \citep{ImbensRubinBook} that depends on eight factors (Table \ref{tabsim}). The distance between treated groups, $b$, is defined in terms of standardized bias $B$, where
\begin{eqnarray} B = \frac{b}{\sqrt{\frac{1+\sigma_2^2+\sigma_3^2}{3}}}\end{eqnarray}

\noindent in order to evaluate the reduction in initial bias somewhat independently of the variance ratios $\sigma_2^2$ and $\sigma_3^2$. 

\begin{table} \centering 
\caption {Simulation factors}
\label{tabsim}
\begin{tabular}{l l}
\hline 
Factor & Levels of factor\\ \hline 
$n_{t_1}$ & $\left \{500,2000 \right\}$ \\
$\gamma=\frac{n_{t_2}}{n_{t_1}} = \frac{n_{t_3}}{n_{t_2}} $ & $\left \{1,2\right\} $\\
$f$& $\left \{t_7, Normal\right\}$\\
$b$ & $B = \frac{b}{\sqrt{\frac{1+\sigma_2^2+\sigma_3^2}{3}}}$ takes levels $\left \{0,0.25,0.50,0.75,1.00 \right\}$ \\
$\tau$ & $\left \{0,0.25 \right\} $\\
$\sigma_2^2$ & $\left \{0.5, 1, 2 \right\}$ \\
$\sigma_3^2$ & $\left \{0.5, 1, 2\right\}$ \\
$P$ & $\left \{3,6\right\}$\\
\hline 
\end{tabular}
\end{table}

Due to the small number of eligible subjects remaining when $P$ = 6 and $n_{t_1}$ = 500, these simulations are discarded, leaving 1080 simulation configurations. For each simulation condition, 200 data sets are generated, and on each data set, $VM$ (using $K$ = 5 strata), $CRM$, $IPW$ and $KMC$ are used to identify matched, weighted, or subclassified sets. For $CRM$, we used $\epsilon$ = 0.25 \citep{austin2011optimal}. 

\subsection{Simulation metrics}
While several metrics have been proposed for evaluating the success of matching with binary treatments (see \citet{austin2007comparison, austin2009balance}, for example), assessments for multiple treatments are not as well formalized \citep{stuart2010matching}. 

For $VM$ or $CRM$, let $n_{trip}$ be the number of triplets formed, and let $\psi_i$ be the number of times subject $i$ is part of a triplet. The weighted mean of covariate $p$, $p = 1, ... P$, at treatment $t$, is defined as $\bar{X}_{pt}$, such that
\begin{eqnarray} \bar{X}_{pt} = \frac{ \sum_{i=1}^N X_{pi} I_i(t) \psi_{i}}{n_{trip}}. \end{eqnarray}

\noindent For $IPW$, $\psi_{i} =\frac{1}{r(t,\boldsymbol{X_i})}$ is each subject's weight, where $r(t,\boldsymbol{X})$ is estimated using multinomial logistic regression, and $n_{trip}$ is simply the number of matched subjects receiving each treatment $t$. With $KMC$, $\bar{X}_{pt}$'s are calculated within each subclass, and weighted across subclasses, with weights proportional to the number of subjects in each subclass. 

For a binary treatment, \citet{rubin1996matching} and \citet{rubin2001using} suggest that the standardized bias between $\boldsymbol{X_p}$ in the treatment $(t_1)$ and control groups $(t_2)$, $SB_{p12}$, should be less than 0.25 to make defensible causal statements, where 
\begin{eqnarray} SB_{p12}  =  \frac{\bar{X}_{p1}- \bar{X}_{p2}}{\delta_{p1}}\end{eqnarray}

\noindent and $\delta_{p1}$ is the standard deviation of $\boldsymbol{X_p}$ in $t_1$.

In our simulations, we calculated three such biases for each covariate $p$ for each pair of treatments, $SB_{p12}$, $SB_{p13}$, and $SB_{p23}$. As in \citet{hade2012propensity}, we extract the maximum absolute standardized pairwise bias at each covariate, $Max2SB_p$, such that
 \begin{eqnarray} Max2SB_{p} = max(|SB_{p12}|,|SB_{p13}|,|SB_{p23}|). \label{2way} \end{eqnarray}

\noindent For all of the matching algorithms and at each $p$, $ \delta_{p1} $, the standard deviation of $\boldsymbol{X_p}$ in the full sample among those receiving reference $t_1$, is used for standardization, to ensure that observed differences in the similarity of those matched are easily contrasted (as in \citet{stuart2008best}).

With three treatment pairs, $Max2SB_{p}$ reflects the largest discrepancy in estimated covariate means between any two treatment groups for a specific covariate. Using a similar metric to assess covariate balance, \citet{mccaffrey2013tutorial} advocated using a standardized bias cutoff of 0.20 for multiple treatments. We also examined average absolute standardized biases, $\frac{|SB_{p12}|+|SB_{p13}|+|SB_{p23}|}{3}$, finding similar results to those with $Max2SB_{p}$. 

In addition to bias, for $VM$ and $CRM$ we also estimated the fraction of units from the entire population who received $t_1$ and were eligible to receive the other two treatments which were included in the final matched set, $\% Matched$.  This metric provides a sense of the similarity between those matched and the population that we are interested in generalizing to. Simulations with $\% Matched \approx 1$ and relatively low $Max2SB_{p} \ \forall \text{ } p$ are optimal in the sense that almost all subjects who received $t_1$ are matched with subjects receiving $t_2$ and $t_3$ and the distributions of their covariates are similar.  $\% Matched$ is not relevant for $IPW$, because weights are estimated for all subjects that meet the eligibility criteria. 

At each simulation configuration and for each each of the matching algorithms, $Max2SB_{p} \text{ } \forall \text{ } p $ and $\% Matched$ are obtained, and averaged across 200 replications. For simplicity, we summarize $Max2SB_{p} \text{ } \forall \text{ } p $ by averaging over $p$, such that $\overline{Max2SB} = \sum_{p=1, ... P} Max2SB_{p}/P$. 

\subsection{Determinants of matching performance}

\begin{figure}[htbp]
\begin{center}
  \includegraphics[width=5in,height=3.33in]{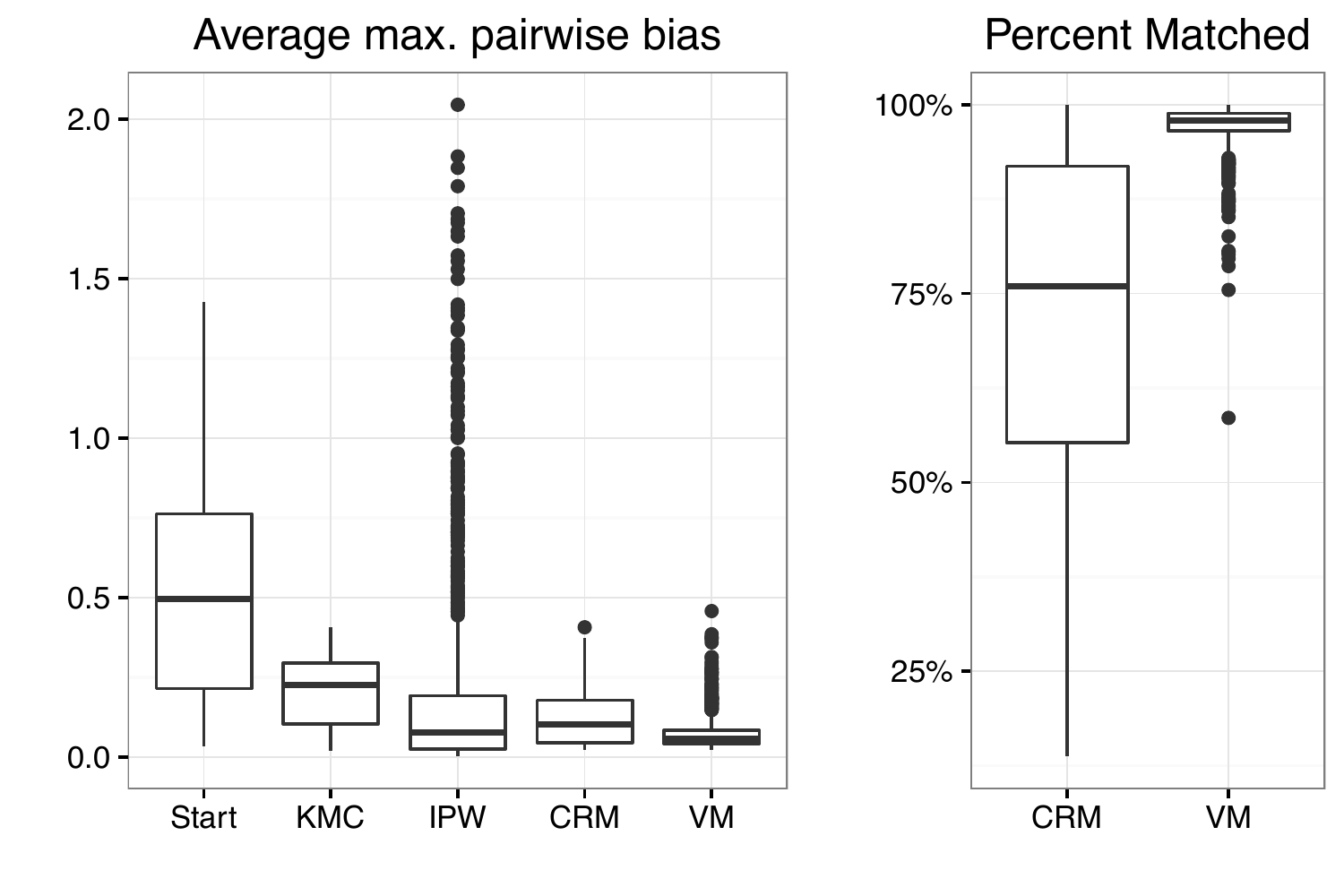}
\caption{$\overline{Max2SB}$ for pre-matched cohort and by matching algorithms (left), and $\%Matched$ for $VM$ and $CRM$}
  \label{outGraphs22}
\end{center}
\end{figure}

Figure \ref{outGraphs22} shows boxplots of $\overline{Max2SB}$ and $\%Matched$ across each of the simulation factors. $\overline{Max2SB}$ was calculated for $VM$, $CRM$, $IPW$, $KMC$, and in the pre-matched cohort of eligible subjects. Each point in each of the boxplots represents the bias at one factors' configuration. In Figure \ref{outGraphs22}, $\overline{Max2SB}$ exceeds a cutoff of 0.20 in 57\% of combinations when using $KMC$, compared to 25\% when using $IPW$, 19\% when using $CRM$ and to 4\% when using $VM$. There are 16 simulation configurations for which $IPW$ yields a $\overline{Max2SB}$ greater than 1.5. In general, $KMC$ has done the worst, with $\overline{Max2SB}$ in more than 75\% of configurations lying above the median $\overline{Max2SB}$ for each of the other algorithms. This corresponds to results for a binary treatment assignment which suggest that subclassification alone may not sufficiently to reduce bias in the covariates' distributions\citep{gutman2013robust}, as well as the problem that some clusters may not include units from all treatment groups. Given its poor performance, $KMC$ is not shown in the tables below. 

$VM$ matched at least 85\% of eligible reference subjects in a matched triplet in 99\% of the configurations, while only 37\% of the configurations for $CRM$ reached the 85\% cutoff. $VM$ matched at least 95\% of the eligible reference group subjects on more than 85\% of the configurations. 

To identify factors with the largest influence on the performance of using $VM$, $CRM$, and $IPW$, we rank them by their MSE for both $\overline{Max2SB}$ as well as $\% Matched$ (as in \cite{rubin1979using}, \cite{cangul2009testing}). Because $\% Matched$  was highly skewed, we used the Box-Cox power transformation \citep{sakia1992box} to make this metric approximately normally distributed. 

Initial covariate bias $B$ drives the highest proportion of variation in $\overline{Max2SB}$, accounting for roughly 85\%, 70\%, and 45\% of the variability for $CRM$, $VM$, and $IPW$, respectively (Table \ref{ANOVA}). Compared to $VM$ and $CRM$, $IPW$ biases' are substantially driven by the distribution type ($f$) and the variance terms $\sigma_2$ and $\sigma_3$. While $\gamma$, the rate of those receiving $t_2$ and $t_3$ relative to the number of subjects receiving $t_1$, is not an important factor for $IPW$, it is the second and third most important factors of $CRM$ and $VM$, respectively. This is also noticed with matching methods for binary treatment \citep{rubin1973matching}. $B$ also drives nearly 100\% of the variability in $\% Matched$ for $VM$ and $CRM$ (not shown). The second most influential factor for the ANOVA of $\% Matched$ using those matched via $CRM$ is $\gamma$; for a binary matching approach, the increased number of available matches on $t_2$ and $t_3$ increases the likelihood that a subject receiving $t_1$ is matched. 

\begin{table} \centering \small
\caption {ANOVA for $VM$, $CRM$ and $IPW$ $\overline{Max2SB}$: most influential factors}
\label{ANOVA}
\begin{tabular}{c c c c c c c c}
\hline 
 \multicolumn{2}{c}{$VM$}  &  &\multicolumn{2}{c}{$IPW$} &  &\multicolumn{2}{c}{$CRM$} \\ \cline{1-2} \cline{4-5} \cline{7-8}
Variable 			& MSE & 		&  Variable & MSE& &  Variable & MSE  \\ \cline{1-2} \cline{4-5} \cline{7-8} 
$B$				&16046 		& 	&$B$		&154397& 		&$B$			&41354\\
$p$				&3776& 		&$f$			&115220& 		&$\gamma$		&4089\\
$\gamma$		&1259& 		&$B$*$f$		&42628& 			&$\sigma_t$		&2271\\
$\sigma_t$		&934& 		&$\sigma_s$	&22903& 			&$\sigma_s$		&469\\
$\sigma_s$		&791& 		&$p$			&14746& 			&$B$:$\sigma_t$	&243\\
$\tau$			&692& 		&$n_{t_1}$	&11707& 			&$B$:$\gamma$	&233\\
$B$*$p$			&692& 		&$\sigma_t$	&10137& 			&$B$*$\sigma_s$	&147\\
$B$*$\sigma_t$	&223& 		&$B$*$\sigma_s$&5280& 		&$p$				&64\\
$p$*$\sigma_s$	&220& 		&$B$*$p$		&3838& 			&$B$*$n_{t_1}$	&48\\
$p$*$\sigma_t$		&216& 		&$B$*$n_{t_1}$&3776& 			&$f$				&28\\
 \hline \hline
\end{tabular}
\end{table}


Having identified the principal determinants of bias and matching size, we average over the other factors in order to further detail the effects of the principal ones. Tables \ref{TabSim1} to \ref{TabSim4} show $\overline{Max2SB}$ based on different biases ($B$), distributions of $\boldsymbol{X}$ ($f$), number of parameters ($P$), number of subjects receiving $t_1$ ($n_{t_1}$), and the ratio of units receiving $t_2$ to those receiving $t_1$ ($\gamma$).

In settings with low bias and normally distributed covariates, all three matching approaches appear to properly balance covariates. The average $\overline{Max2SB}$ using $IPW$ is less than 0.05 across each simulation configuration with $B=0$ and $f$ = Normal. As $B$ increases, $\overline{Max2SB}$ for $CRM$ rises faster than for $VM$. $IPW$ bias also rises with higher $B$, but in most settings with normally distributed covariates, $IPW$ yields $\overline{Max2SB}$ less than 0.25, but higher than $VM$. 

$VM$ and $CRM$ produce better matched groups than $IPW$ with heavy tailed covariates. When the covariates are distributed as multivariate $t_7$, the maximum pairwise bias's using $IPW$ vary substantially (e.g., Table \ref{TabSim4}). While $\gamma$ is not a major determinant of $\overline{Max2SB}$ for $IPW$, $VM$ and $CRM$ perform better in settings with $\gamma=2$ (Tables \ref{TabSim2} and \ref{TabSim4}). 

Table \ref{TabSipm} shows the $\% Matched$ for different values of $n_{t_1}$ and $\gamma$, averaging over $P$, $f$, $\tau$, $\sigma_2$ and $\sigma_3$. For low bias and with a larger number of controls ($\gamma = 2$), $CRM$ generally matches as many triplets as $VM$. With increasing $B$, however, the fraction of eligible units that were matched is much smaller for $CRM$. With $\gamma=1$, $B=1$, and $n_{t_1}$ = 1000, for example, $CRM$ matches only 36\% of eligible subjects on average, compared to 93\% of the subjects using $VM$.

To account for the smaller number of subjects matched using $VM$, which is a possible unfair advantage for $VM$, we also measured bias in the covariates' distributions for $IPW$ using only the subjects that were utilized by $VM$. In more than 98\% of configurations, the biases observed were larger than those using $IPW$ with all units.

A reduced set of simulations using $Z = 5$ showed that both $VM$ and $IPW$ reduce the initial bias. In some scenarios $VM$ had larger reduction than $IPW$, and in some scenarios the opposite (see Appendix \ref{AppA} for additional details).

\section{Conclusion} \label{Disc}

Many real world problems involve making a decision among three or more possible interventions. Simultaneous assessment of all of these interventions is attractive, because it allows for the identification of the best intervention without the need to perform many studies in which each pair of interventions is compared. However, even in a randomized controlled environment, multi-arm trials can be considerably more complex to design, conduct and analyze than two-arm, single-question trials \citep{vermorken2005clinical}. These complications include sample size requirements, eligibility of all participants for all of the interventions, the comparisons that will be made, as well as the summaries of those comparisons. These problems are exacerbated in non-randomized settings. While estimating causal effects for binary treatment in randomized and non-randomized settings has been discussed extensively in the literature, we highlighted how the specification of causal effects for multiple treatments may be complex due to the choice of estimands and the different subsets of the population which investigators are interested in. Different estimands may yield different conclusions with respect to treatment effectiveness, and we advocate that researchers consider carefully the causal effect, or sets of causal effects, of primary interest, as in \citet{dore2013different}.

\subsection{Discussion of vector matching}

We demonstrated that matching on a vector can address some of the drawbacks of currently available methods for estimating treatment effects with a nominal treatment assignment. $VM$ attempts to replicate a randomized multi-arm trial by generating sets of subjects that are roughly equivalent on measured covariates. Simulations demonstrated that, relative to other available methods, $VM$ generally yielded the lowest bias in the covariates' distributions between the different treatment groups, while retaining most of the eligible subjects that received the reference treatment. Under regular assignment mechanism, differences in $VM$ matched units' outcomes could be contrasted, providing treatment effects that can be generalized to the population of subjects receiving $t_1$. 

$VM$ is a starting point for algorithms that intend to estimate transitive treatment effects and reduce bias when comparing multiple treatments. It is worth explicating on a few of the algorithm's strengths and weaknesses. $VM$ uses with replacement matching because it has been shown to yield lower bias in comparison to matching without replacement with binary treatment \citep{abadie2006large}. Additionally, matching with replacement allows estimation of $PATT$'s which are generalizable to each treatment group, and not just the group with the smallest sample size. One difficulty of matching with replacement is that subjects can be matched multiple times. As a result, although no adjustments are necessary for point estimates, an analysis phase will require adjustments for estimating sampling variances \citep{abadie2006large}. 

$VM$ can be used to estimate any causal estimand of interest and is not restricted to differences in averages. While we concentrated on $PATT$'s in this manuscript, $VM$ can be extended for $PATE$'s by forming a matched set for each eligible unit, as opposed to just a set for each unit receiving the reference treatment. If all eligible subjects can be matched to subjects receiving other treatments, contrasts between the matched cohorts would generalize to the population as a whole. As noted in \citet{abadie2006large}, pair matching for $PATE$'s can only be done with replacement, as differences in the sample sizes at each treatments will require some subjects to be matched more often than others. In this respect, $VM$ would be preferred to $CRM$, $SBC$, and $WithinTrio$, which are limited to only estimating $PATT$'s. 

While $KMC$ is one approach for grouping similar subjects, by restricting the matching to be within the clusters, some possible matches may not be considered by $VM$ because they are on the boundaries of the clusters. This could lead to non-optimal matches, or even to the exclusion of some reference units that will not have a match in the other treatment groups. One plausible extension of $VM$ would be to use fuzzy clustering \citep{bezdek1984fcm}, which would allow for units to belong to multiple clusters.

Another downside of $KMC$ is the possibility of obtaining clusters where there are no units receiving a certain treatment. However, clustering on $Z-2$ components of the GPS, as in vector matching, is preferred to clustering on all $Z$ components, as would be done in using $KMC$ alone. For large $Z$, if clustering on $Z-2$ components yields clusters without at least one unit from each treatment group, one possibility is to re-fit $KMC$, given that $K-means$ often returns different partitions.

Finally, $VM$ is based on a greedy matching algorithm, which may not be the most optimal procedure to partition the GPS. Among other alternatives to matching on the GPS, coarsened exact matching could be used to pair subjects within each of the $K-means$ subclasses. \citep{iacus2011causal}. With binary treatment, algorithms like full matching \citep{rosenbaum1991characterization} and mixed integer programming \citep{zubizarreta2012using} were proposed to optimally match units such that the difference in the covariates' distributions between the two treatment groups is minimized while retaining most of the units. In contrast to binary treatment matching, optimally matching for multiple treatments, also known as $k$-dimensional matching, was shown to be a NP-hard problem \citep{karp1972reducibility}. Further research is required to apply these methods to multiple treatments.

As with other procedures for estimating causal effects with multiple treatments, methods for estimating the sampling variances of estimands when using $VM$ are not well established. Variance weighting and resampling are two procedures that have been suggested for estimating the sampling variance of causal estimands with binary treatments, and we proposed that similar procedures could be used with multiple treatments. However, further research is required to identify the operating characteristics of each of these procedures.

One set of strategies that we did not explore is covariate adjustment for the GPS or a function of the GPS using a regression model \citep{filardo2009relation, filardo2007obesity, dearing2009does, spreeuwenberg2010multiple}. Such techniques are subject to possible model misspecification and extrapolation problems, as shown in standard regression adjustment for binary treatment \citep{dehejia1998causal, dehejia2002propensity}, and simulations have found that these strategies can perform worse than matching, stratification, or weighting with multiple treatments \citep{hade2013bias}.

\subsection{Recommendations}

Causal modeling is challenging because it requires estimation of quantities that cannot be measured simultaneously. This problem is exacerbated when comparing multiple treatments, because the proportion of these quantities increases. Methods for multiple treatments continue to evolve, and more work is still needed in several areas, particularly with respect to the estimation of the sampling variance. Below, we provide a list of recommendations for researchers who are looking to estimate causal effects with multiple treatments.

\begin{enumerate}

\item Comparing multiple treatments in observational studies is similar to comparing multiple interventions in a multi-arm trial. Thus, it is important to ascertain that the data is composed of enough units that are `eligible' to receive all of the treatments, and units that are not eligible should be removed when attempting to identify the best treatment.

\item Causal estimands of interest and the populations to which these estimands generalize require careful consideration. These decisions become more complex with increased number of treatments.

\item For ordinal treatment assignment such as scales or doses, the linear predictor from a proportional odds model of treatment assignment acts as a scalar balancing score on which to balance the covariates' distributions. Non-bipartite matching \citep{lu2001matching}, subclassification \citep{imai2004causal, zanutto2005using}, and the combination of subclassification with regression adjustment \citep{lopez2014estimating} stand out as approaches for making inferences. 

\item For nominal treatment assignment, methods that rely on binary propensity scores that are estimated only on units receiving one of two treatments (such as $SBC$ and $CRM$) may result in significant bias in the covariates' distributions between units receiving the different treatments. These may lead to biased and non-transitive estimates, and therefore should not be applied generally.

\item  For nominal or ordinal treatment assignment, a simple implementation of $K-means$ clustering ($KMC$) may result in clusters that do not include units receiving all treatments, which results in increased bias. Our simulations show that in comparison to other matching and weighting procedures, it suffers from the smallest bias reduction.

\item For nominal or ordinal treatment assignment, matching on the GPS using vector matching ($VM$) or using inverse probability weighting ($IPW$) are promising approaches.

\begin{itemize} 

\item $IPW$ reduces the bias significantly; however, as our simulations show, it may suffer from extreme weights that yield erratic causal estimates. This problem is exacerbated with increasing number of treatments or covariates that are not normally distributed. Simple trimming of units with GPS components that are close to 0 or 1 may result in increased bias, because units that are similar on a single GPS component may differ on others. Other approaches for estimating the GPS, such as generalized boosted models, may solve this issue. However, more research is needed to derive sampling variance estimates for these procedures and to examine their behavior in a wide range of applications. Lastly, $IPW$ estimates are mainly suitable for estimating differences in averages, and are not well suited for comparison of other estimands.

\item $VM$ uses an in-strata matching algorithm to identify matched sets of subjects in order to estimate treatment effects generalizable to the population of units eligible for each treatment. Across a set of simulation configurations, $VM$ tended to yield the largest improvement in balance in the covariates' distributions between units receiving different treatments. Under certain assumptions, this would allow for unbiased comparisons of the effects of multiple interventions. Additional research is needed to identify sampling variance formulas for estimates from the matched cohorts, as well as to explore alternative mechanisms for matching on the GPS.  To sum, $VM$ is one approach that seems to compare favorably to commonly available methods, but more research is needed to explore other alternatives.

\end{itemize}

\end{enumerate}

\clearpage

\begin{table}[ht] \centering \small 
\caption {$\overline{Max2SB}$, small/equal sample sizes: $n_{t_1} = 500$, $n_{t_2} = 500$, $n_{t_3} = 500$}
\label{TabSim1}
\begin{tabular}{c c c c c c c c c}
\hline 
   & \multicolumn{7}{c}{$P$=3} \\  \cline{2-8} 
   & \multicolumn{3}{c}{$f$ = Normal } &  & \multicolumn{3}{c}{$f$ = $t_7$ }  \\\cline{2-4} \cline{6-8} 
 $B$  & $VM$  &$CRM$& $IPW$ & & $VM$  &$CRM$& $IPW$ \\ \hline
0.00&0.06&0.06&0.01&&0.05&0.06&0.01\\
0.25&0.06&0.08&0.03&&0.06&0.08&0.04\\
0.50&0.07&0.13&0.07&&0.07&0.13&0.11\\
0.75&0.10&0.20&0.11&&0.09&0.19&0.30\\
1.00&0.15&0.25&0.17&&0.14&0.26&0.58\\
\hline 
\end{tabular}
\end{table}
\begin{table}[h] \centering \small
\caption {$\overline{Max2SB}$, small/unequal sample sizes: $n_{t_1} = 500$, $n_{t_2} = 1000$, $n_{t_3} = 2000$}
\label{TabSim2}
\begin{tabular}{c c c c c c c c c}
\hline 
   & \multicolumn{7}{c}{$P$=3} \\  \cline{2-8} 
   & \multicolumn{3}{c}{$f$ = Normal } &  & \multicolumn{3}{c}{$f$ = $t_7$ }  \\\cline{2-4} \cline{6-8} 
 $B$  & $VM$  &$CRM$& $IPW$ & & $VM$  &$CRM$& $IPW$ \\ \hline
0.00&0.04&0.05&0.01&&0.05&0.05&0.01\\
0.25&0.05&0.05&0.04&&0.05&0.05&0.04\\
0.50&0.05&0.08&0.08&&0.06&0.07&0.11\\
0.75&0.07&0.13&0.12&&0.07&0.13&0.29\\
1.00&0.10&0.20&0.17&&0.10&0.19&0.60\\
\hline 
\end{tabular}
\end{table}
\begin{table}[hp] \centering \small
\caption {$\overline{Max2SB}$, large/equal sample sizes: $n_{t_1} = 1000$, $n_{t_2} = 1000$, $n_{t_3} = 1000$}
\label{TabSim3}
\begin{tabular}{c c c c c c c c c c c c c c c c}
\hline 
   & \multicolumn{7}{c}{$P$=3} &  & \multicolumn{7}{c}{$P$=6}\\  \cline{2-8} \cline{10-16}
   & \multicolumn{3}{c}{$f$ = Normal } &  & \multicolumn{3}{c}{$f$ = $t_7$ } &  & \multicolumn{3}{c}{ $f$ = Normal } &  & \multicolumn{3}{c}{$f$ = $t_7$  }    \\\cline{2-4} \cline{6-8} \cline{10-12} \cline{14-16}
 $B$  & $VM$  &$CRM$& $IPW$ & & $VM$  &$CRM$& $IPW$ & &$VM$  &$CRM$& $IPW$ & &$VM$  &$CRM$& $IPW$  \\ \hline
0.00&0.04&0.03&0.01&&0.04&0.04&0.01&&0.05&0.04&0.01&&0.05&0.04&0.01\\
0.25&0.04&0.07&0.03&&0.04&0.07&0.04&&0.05&0.08&0.04&&0.05&0.07&0.05\\
0.50&0.05&0.14&0.07&&0.05&0.13&0.11&&0.09&0.14&0.08&&0.08&0.14&0.21\\
0.75&0.07&0.20&0.11&&0.07&0.20&0.35&&0.13&0.19&0.14&&0.14&0.19&0.66\\
1.00&0.10&0.26&0.16&&0.10&0.25&0.75&&0.23&0.24&0.23&&0.24&0.26&1.06\\
\hline 
\end{tabular}
\end{table}

\begin{table}[hp] \centering \small 
\caption {$\overline{Max2SB}$, large/equal sample sizes: $n_{t_1} = 1000$, $n_{t_2} = 2000$, $n_{t_3} = 4000$}
\label{TabSim4}
\begin{tabular}{c c c c c c c c c c c c c c c c}
\hline 
   & \multicolumn{7}{c}{$P$=3} &  & \multicolumn{7}{c}{$P$=6}\\  \cline{2-8} \cline{10-16}
   & \multicolumn{3}{c}{$f$ = Normal } &  & \multicolumn{3}{c}{$f$ = $t_7$ } &  & \multicolumn{3}{c}{ $f$ = Normal } &  & \multicolumn{3}{c}{$f$ = $t_7$  }    \\\cline{2-4} \cline{6-8} \cline{10-12} \cline{14-16}
 $B$  & $VM$  &$CRM$& $IPW$ & & $VM$  &$CRM$& $IPW$ & &$VM$  &$CRM$& $IPW$ & &$VM$  &$CRM$& $IPW$  \\ \hline
0.00&0.03&0.03&0.01&&0.03&0.03&0.01&&0.04&0.04&0.01&&0.04&0.04&0.01\\
0.25&0.03&0.04&0.04&&0.03&0.03&0.04&&0.04&0.04&0.04&&0.05&0.04&0.05\\
0.50&0.04&0.08&0.08&&0.04&0.07&0.12&&0.06&0.09&0.09&&0.06&0.08&0.20\\
0.75&0.05&0.14&0.13&&0.05&0.12&0.33&&0.11&0.16&0.14&&0.09&0.15&0.67\\
1.00&0.08&0.21&0.17&&0.07&0.19&0.80&&0.17&0.21&0.22&&0.16&0.21&1.28\\
\hline 
\end{tabular}
\end{table}

\begin{table} [h]\centering \small 
\caption {$\% Matched$: The percent of eligible subjects receiving $t_1$ who were matched}
\label{TabSipm}
\begin{tabular}{c c c c c c c c c c c c}
\hline 
   & \multicolumn{5}{c}{$n_{t_1}=500$} &  & \multicolumn{5}{c}{$n_{t_1}$=1000}\\  \cline{2-6} \cline{8-12}
   & \multicolumn{2}{c}{$\gamma$ = 1 } &  & \multicolumn{2}{c}{$\gamma$ = 2  } &  & \multicolumn{2}{c}{ $\gamma$ = 1 } &  & \multicolumn{2}{c}{$\gamma$ = 2   }    \\\cline{2-3} \cline{5-6} \cline{8-9} \cline{11-12}
 $B$  & $VM$  &$CRM$&  & $VM$  &$CRM$& &$VM$  &$CRM$&  &$VM$  &$CRM$  \\ \hline
0.00&0.99&0.91&&0.99&0.99&&0.99&0.92&&0.99&0.99\\
0.25&0.97&0.81&&0.99&0.97&&0.98&0.79&&0.99&0.96\\
0.50&0.95&0.67&&0.98&0.87&&0.98&0.63&&0.99&0.79\\
0.75&0.94&0.52&&0.97&0.72&&0.97&0.47&&0.98&0.61\\
1.00&0.91&0.42&&0.95&0.55&&0.93&0.36&&0.96&0.47\\
\hline 
\end{tabular}
\end{table}

\clearpage

\begin{table}[ht] \centering \small 
\caption {Summary of acronyms}
\label{TabAcr}
\begin{tabular}{c c c c c c c c c}
\hline 
Acronym & Description \\ \hline
$CRM$ & Common referent matching \\
GPS & Generalized propensity score \\
$IPW$ & Inverse probability weighting\\
$KMC$ & K-means clustering\\
$PATE$ & Population average treatment effects\\
$PATT$ & Population average treatment effects among the treated\\
RCM & Rubin causal model\\
$SATE$ & Sample average treatment effects\\
$SATT$ & Sample average treatment effects among the treated\\
$SBC$ &Series of binary comparisons\\
$SUTVA$ & Stable unit treatment value assumption \\
$VM$ & Vector matching\\
\hline 
\end{tabular}
\end{table}

\clearpage

\section{Appendix} \label{AppA}

We implement $VM$ and $IPW$ for $Z=5$, where $\boldsymbol{X}$ is generated for $N = n_{t_1} + n_{t_2} +n_{t_3} + n_{t_4} +n_{t_5}$ subjects receiving one of five treatments, $\mathcal{T} \in \left\{t_1,t_2,t_3,t_4,t_5\right\}$, with $n_{t}$ the sample size of subjects receiving treatment $t$. Let $\mathbb{1}$ be the 5 x 5 identify matrix. The values of $\boldsymbol{X}$ were generated from multivariate symmetric distributions such that
\begin{eqnarray} T_i &=& t_1, i = 1, ...,n_{t_1}\\
 	 T_i &=&t_2, i = n_{t_1}+1, ..., n_{t_1}+\gamma n_{t_1}  \nonumber \\
	  T_i &=& t_3, i = n_{t_1}+\gamma n_{t_1}+1, ..., n_{t_1}+2*\gamma n_{t_1}\nonumber \\
 	 T_i &=&t_4, i = n_{t_1}+2*\gamma n_{t_1}+1, ..., n_{t_1}+2*\gamma n_{t_1} + \gamma^2*n_{t_1} \nonumber \\
	  T_i &=& t_5, i = n_{t_1}+2*\gamma n_{t_1} + \gamma^2*n_{t_1} +1, ..., n_{t_1}+2*\gamma n_{t_1} + 2*\gamma^2*n_{t_1}   \nonumber  \end{eqnarray}
\begin{eqnarray}
	\boldsymbol{X_i} |\left\{T_i=t_1\right\}& \sim& f(\boldsymbol{\mu_1}, \boldsymbol{\Upsigma}),i = 1, ...,n_{t_1}\\
	\boldsymbol{X_i} |\left\{T_i=t_2\right\} &\sim& f(\boldsymbol{\mu_2}, \boldsymbol{\Upsigma}), i = n_{t_1}+1, ..., n_{t_1}+\gamma n_{t_1}  \nonumber\\
	\boldsymbol{X_i} |\left\{T_i=t_3\right\} &\sim& f(\boldsymbol{\mu_3}, \boldsymbol{\Upsigma}), i = n_{t_1}+\gamma n_{t_1}+1, ..., n_{t_1}+2*\gamma n_{t_1} \nonumber \\ 
	\boldsymbol{X_i} |\left\{T_i=t_4\right\} &\sim& f(\boldsymbol{\mu_4}, \boldsymbol{\Upsigma}), i = n_{t_1}+2*\gamma n_{t_1}+1, ..., n_{t_1}+2*\gamma n_{t_1} + \gamma^2*n_{t_1}  \nonumber \\ 
	\boldsymbol{X_i} |\left\{T_i=t_5\right\} &\sim& f(\boldsymbol{\mu_5}, \boldsymbol{\Upsigma}),  i = n_{t_1}+2*\gamma n_{t_1} + \gamma^2*n_{t_1} +1, ..., n_{t_1}+2*\gamma n_{t_1} + 2*\gamma^2*n_{t_1} \nonumber \end{eqnarray}
\begin{eqnarray}	\mu_1= (b,0,0,0,0)^T, \mu_2 = (0,b,0,0,0)^T,\mu_3= (0,0,b,0,0)^T, \mu_4 = (0,0,0,b,0)^T,\mu_5= (0,0,0,0,b)^T \end{eqnarray}
\begin{eqnarray}	\Upsigma = \mathbb{1} \end{eqnarray}

The following design implicitly assumes a regular assignment mechanism that depends on four factors (Table \ref{tabsimnew}). 

\begin{table}[h] \centering 
\caption {Simulation factors}
\label{tabsimnew}
\begin{tabular}{l l}
\hline 
Factor & Levels of factor\\ \hline 
$n_{t_1}$ & $\left \{1000 \right\}$ \\
$\gamma=\frac{n_{t_2}}{n_{t_1}} = \frac{n_{t_3}}{n_{t_1}}=\frac{n_{t_4}}{n_{t_2}} = \frac{n_{t_5}}{n_{t_2}} $ & $\left \{1,2\right\} $\\
$f$& $\left \{t_7, Normal\right\}$\\
$b$ & $\left \{0,0.25,0.50,0.75,1.00 \right\}$ \\
\hline 
\end{tabular}
\end{table}

For each simulation condition, 200 data sets are generated, and on each data set, $VM$ (using $K$ = 5 strata) and $IPW$ are used to identify matched and weighted sets. $CRM$ is not considered do to the small number of matches generated. 

In all 20 configurations, both $VM$ and $IPW$ reduced the average $\overline{Max2SB}$ relative to the pre-matched cohort. In cases with large initial bias and with covariates from $t$ distribution, $VM$ performed better than $IPW$, but with smaller initial bias, $IPW$ performed better. On average, $VM$ matched at least 93\% of eligible subjects in each configuration. Investigating the performance of different matching methods with five or more treatments is an area of further research.

\bibliographystyle{imsart-nameyear}
\bibliography{ReferencesIII}

\end{document}